\documentclass[a4paper,11pt,fleqn]{article}
\pdfoutput=1
\usepackage{jinstpub}
\usepackage{array}
\usepackage{subcaption}
\usepackage{hyperref}
\usepackage[mathlines]{lineno}

\usepackage{xcolor}

\newcolumntype{A}{>{\raggedright\arraybackslash}m{9cm}}

\title{Estimation of Combinatoric Background in SeaQuest using an Event-Mixing Method}
\author[a,1]{S.F. Pate,\note{Corresponding Author}}
\author[a]{A.~Pun,}
\author[a]{M.F.~Hossain,}
\author[b]{K.~Nagai,}
\author[c]{C.A.~Aidala,}
\author[c]{C.~Ayuso,}
\author[d,e]{L.~El Fassi,}
\author[f]{D.F.~Geesaman,}
\author[g,2]{T.J.~Hague,\note{currently at Lawrence Berkeley National Laboratory, Berkeley, CA, 94720, USA}}
\author[h]{E.R.~Kinney,}
\author[c]{W.~Lorenzon,}
\author[i,j]{K.~Nakano,}
\author[f]{P.E.~Reimer,}
\author[c]{M.B.C.~Scott,}
\author[g]{R.S.~Towell}
\affiliation[a]{Department of Physics, New Mexico State University, Las Cruces, NM, 88003, USA}
\affiliation[b]{Physics Division, Los Alamos National Laboratory, Los Alamos, NM, 87545, USA}
\affiliation[c]{Department of Physics, University of Michigan, Ann Arbor, MI, 48109, USA}
\affiliation[d]{Department of Physics and Astronomy, Rutgers, The State University of New Jersey, Piscataway, NJ, 08854, USA}
\affiliation[e]{Department of Physics and Astronomy, Mississippi State University, Mississippi State, MS, 39762, USA}
\affiliation[f]{Physics Division, Argonne National Laboratory, Lemont, IL, 60439, USA}
\emailAdd{spate@nmsu.edu}
\affiliation[g]{Department of Engineering and Physics, Abilene Christian University, Abilene, TX, 79601, USA}
\affiliation[h]{Department of Physics, University of Colorado, Boulder, CO, 80309, USA}
\affiliation[i]{Department of Physics, University of Virginia, Charlottesville, VA, 22904, USA}
\affiliation[j]{RIKEN Nishina Center for Accelerator-Based Science, Wako, Saitama
351-0198, Japan}

\abstract{
All experiments observing dilepton pairs (e.g.\ $e^+e^-$, $\mu^+\mu^-$) must confront the existence of a {\em combinatoric} background caused by the combining of tracks not arising from the same physics vertex.  Some method must be devised to calculate and remove this background.  In this document we describe a particular event-mixing method relying on many of the unique aspects of the SeaQuest spectrometer and data.  The method described here calculates the combinatoric background with correct normalization; i.e., there is no need to assign a floating normalization factor that is then determined in a subsequent fitting procedure.  Numerous tests are applied to demonstrate the reliability of the method.} 


\begin{document}

\maketitle

\section{Introduction}
The SeaQuest experiment looks for dimuon signals coming from either the Drell-Yan process or from the decay of the $J/\psi$ or $\psi'$ mesons~\cite{SeaQuest:2017kjt,Dove_2021}. The track pairs we reconstruct from the data not only contain muon pairs from the aforementioned processes but also random combinations of single muons from uncorrelated processes. These background pairs arise not only from multiple physics interactions in the same beam bunch, but also from complex single events like open-charm production, where each charmed meson decays into a muonic channel. Such random combinations of muons, called the {\em combinatoric} background, need to be subtracted from the data to extract the true signal yields. We discuss an event-mixing method to estimate the combinatoric background from the SeaQuest data, and we demonstrate that this method has the correct absolute normalization.\footnote{This manuscript does {\em not} describe the method used in Ref.~\cite{Dove_2021} to estimate the combinatoric background.  The method described here has been developed later.}

Crochet and Braun-Munzinger~\cite{Crochet:2001qd} emphasize two important characteristics of a successful event-mixing process for estimating the combinatoric background.
Firstly, the signal density in the data stream must be low. This is the case in SeaQuest, where only approximately 5\% of events in the data stream contain a candidate dimuon pair from a photon or vector meson decay; the vast majority of triggered events do not contain tracks from a physics signal.  (Additional analysis requirements on these candidate tracks will reduce this raw 5\% value.)
Secondly, the tracks must be mixed from events that are ``similar" to each other. This "similarity" is to guarantee that the tracks are drawn from events with similar track distributions.  In Ref.~\cite{Crochet:2001qd} the discussion is concentrated around heavy-ion collision data, so it is suggested to divide the events up into centrality and flow classes; centrality strongly affects track multiplicity and flow introduces momentum correlations.  In SeaQuest, we will see that the relevant quantity is the station-1 drift chamber occupancy, which is largely driven by the tremendous variation in proton beam bunch sizes delivered to the target; therefore we will sort events by the chamber occupancy.

\section{Characteristics of the SeaQuest Data Sample}

\begin{figure}[ht]
\centering
\includegraphics[width=0.9\linewidth]{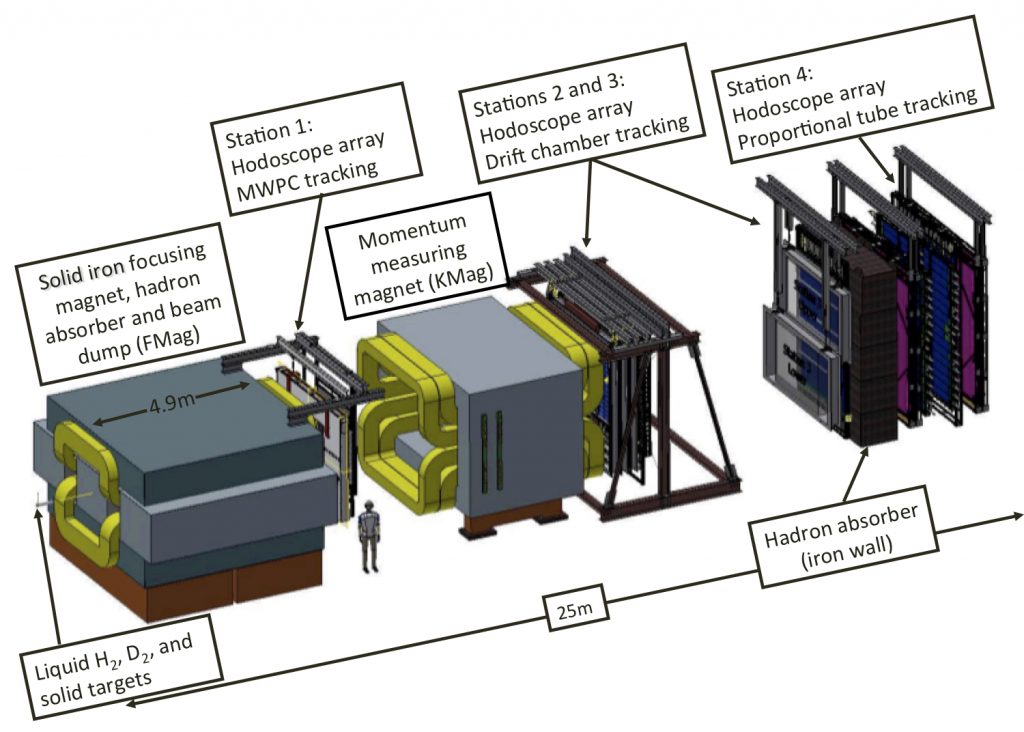}
\caption{Perspective view of the SeaQuest spectrometer.}
\label{fig:spec}
\end{figure}

The SeaQuest spectrometer is fully described in Ref.~\cite{SeaQuest:2017kjt} and a schematic diagram is shown in Fig.~\ref{fig:spec}. A 120 GeV proton beam from the Fermilab Main Injector was incident upon liquid hydrogen and deuterium targets, and also on variety of solid targets.  Particles produced by interactions in the target passed into a 5m-thick iron beam dump, which served to absorb all particles except for highly energetic muons.  The beam dump was also a magnet which served to focus the muons into the spectrometer.  Following the beam dump were the ``station 1'' detectors, comprising $x$- and $y$-measuring hodoscopes and six planes of drift chambers.  Following station 1 was an open magnet which served to measure the momentum of the muons.  Following this magnet came stations 2 and 3, each composed of hodoscopes and wire chambers.  After station 3 was an additional iron absorber for muon identification purposes.  Lastly came station 4 comprising hodoscopes and proportional tubes.  The hodoscopes were used for triggering purposes, while the wire chambers and proportional tubes were used for track reconstruction.  SeaQuest took physics data during the years 2014-2017.

The event trigger in SeaQuest is determined by a field-programmable gate array (FPGA) which looks at the $x$-measuring (bend-plane) hodoscopes at the four stations of the spectrometer.  The FPGA is programmed to look for likely opposite-sign track pairs, based on a simulation of such pairs passing through the hodoscope stations.  Even though the FPGA trigger rejects a tremendous amount of useless particle tracks, it is still programmed to be a ``loose'' trigger, in the sense that it rejects very few valid events.  In this study, we will only use one of the FPGA triggers, the one named ``top/bottom'', which looked for pairs where one track passed through the upper half of the spectrometer and the second track passed through the lower half.  

For the purposes of studying the proposed event-mixing method, we used ten one-hour data runs, from the calendar year 2015 beam period.  The beam quality was typical for SeaQuest runs; the distribution of the number of protons per 1-ns bucket had a mean of about 25,000 accompanied by a long tail extending to approximately 80,000 protons-per-bucket.  The large variation in protons-per-bucket resulted in a large variation in the drift chamber occupancy per event.

\begin{figure}[ht]
\centering
\includegraphics[width=0.5\linewidth]{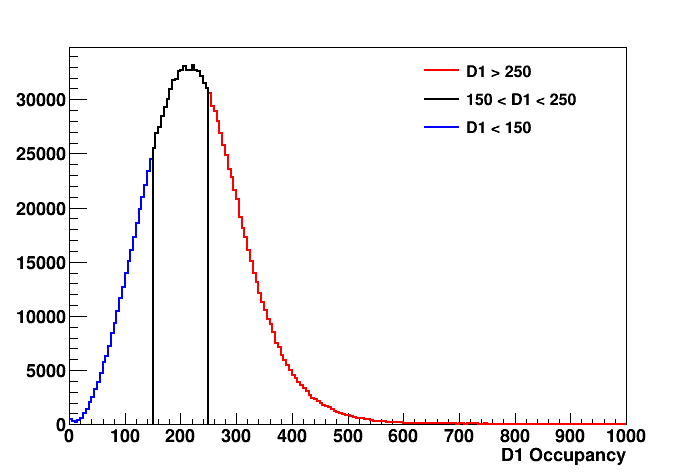}
\caption{Distribution of the D1 occupancy ($\omega$) for top/bottom-triggered events. The vertical lines show the separation between the regions of low ($\omega<150$), middle ($150<\omega<250$) and high ($\omega>250$) occupancy.}
\label{fig:D1}
\end{figure}

The occupancy of the drift chambers at station 1, $\omega$, is defined as the number of hits in the station-1 drift chambers, collectively called D1. The number of wires in each plane in D1 is shown in Table~\ref{tab:wires}.
\begin{table}[ht]
    \centering
    \begin{tabular}{cc}
        \hline
        Plane Name & Number of wires \\
        \hline\hline
        D1U, D1Up, D1V, D1Vp & 201 for each plane \\
        D1X, D1Xp & 160 for each plane \\
        \hline
        Total & 1124 \\
        \hline
    \end{tabular}
    \caption{Number of wires on each plane in the station-1 drift chambers.}\label{tab:wires} 
\end{table}
The D1 occupancy distribution for top/bottom-triggered events is shown in Fig.~\ref{fig:D1}.

\begin{figure}[ht]
\centering
\includegraphics[width=0.45\linewidth]{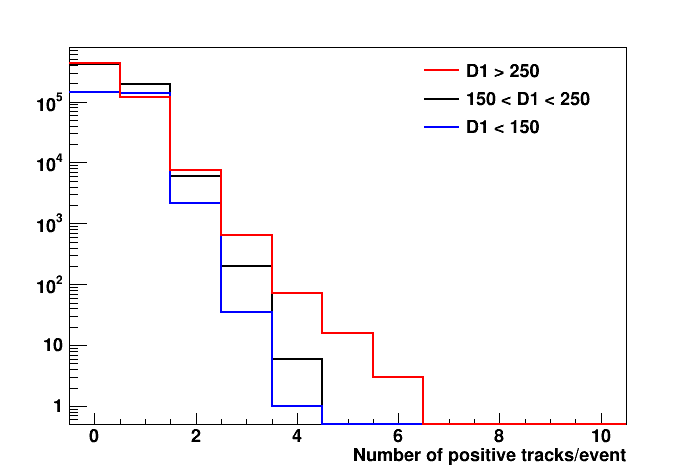}
\includegraphics[width=0.45\linewidth]{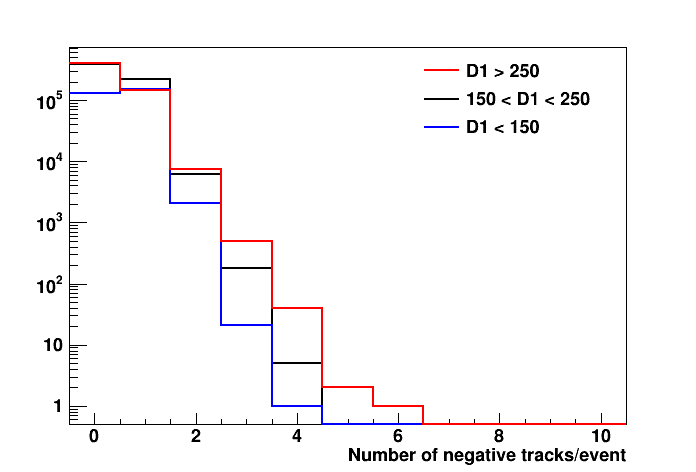}
\caption{Track multiplicities for positive (left) and negative (right) tracks at different D1 occupancies. The blue, black, and red lines correspond to low, middle and high occupancy values, respectively. There are in general more positive tracks than negative tracks because more positive pions than negative pions are produced in collisions of protons with nuclei.}
\label{fig:mult_D1}
\end{figure}

\begin{figure}[ht]
\centering
\includegraphics[width=0.45\linewidth]{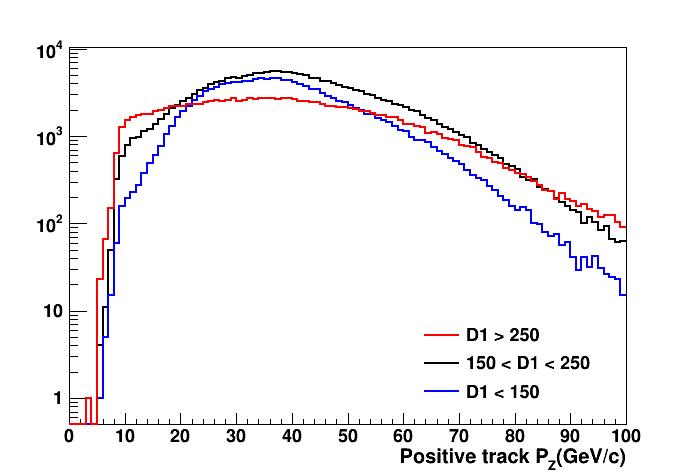}
\includegraphics[width=0.45\linewidth]{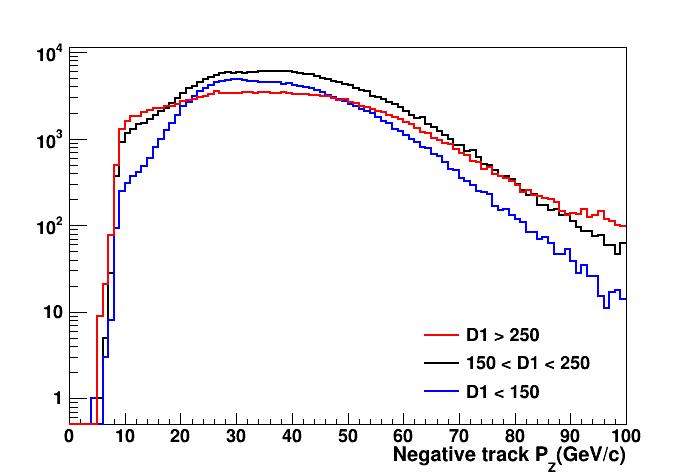}
\caption{Longitudinal momentum $P_z$ for positive (left) and negative (right) tracks at different D1 occupancies. The blue, black, and red lines correspond to low, middle and high occupancy values, respectively.}
\label{fig:pz_D1}
\end{figure}

\begin{figure}[ht]
\centering
\includegraphics[width=0.45\linewidth]{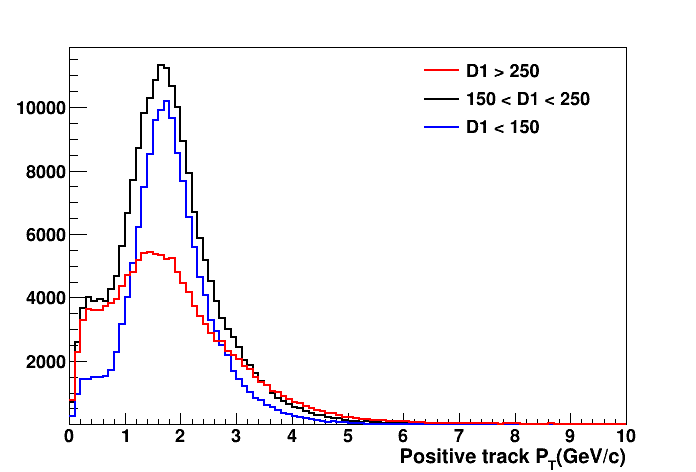}
\includegraphics[width=0.45\linewidth]{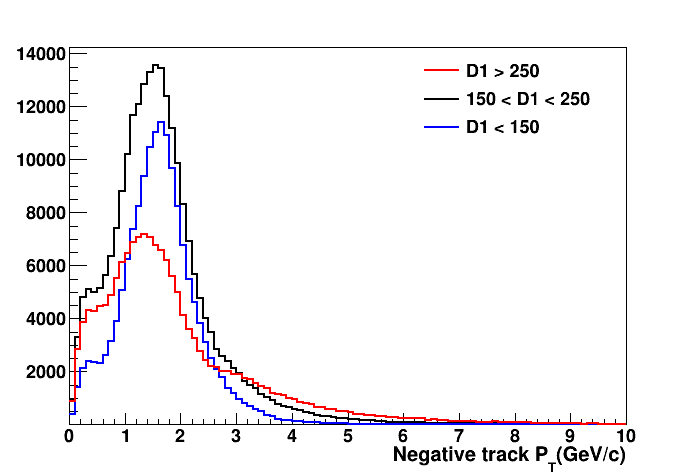}
\caption{Transverse momentum $p_T$ for positive (left) and negative (right) tracks at different D1 occupancies. The blue, black, and red lines correspond to low, middle and high occupancy values, respectively.}
\label{fig:pt_D1}
\end{figure}

The D1 occupancy is divided into three different regions; low ($\omega<150$), middle ($150<\omega<250$) and high ($\omega>250$). Track multiplicities and momentum ($P_z$ and $p_T$) for different D1 occupancy bins are compared. These histograms are produced using the raw reconstructed tracks; no additional cuts have been applied.  The track (positive and negative) multiplicity for different occupancy regions are shown in Fig.~\ref{fig:mult_D1}. The number of tracks per event increases with increasing occupancy. The $P_z$ momentum distributions for both positive and negative tracks for different occupancy regions are shown in Fig.~\ref{fig:pz_D1}. The distribution becomes flatter and wider for higher occupancies. The $p_T$ distributions for both positive and negative tracks for different occupancy regions are shown in Fig.~\ref{fig:pt_D1}. The $p_T$ distribution changes greatly for higher occupancies. 

In summary, the SeaQuest data stream has a number of important features:
\begin{itemize}
    \item The majority of top/bottom-triggered events contain zero reconstructed tracks.  A similarly large fraction contain only one track.
    \item Due to the low density of signal events in the data stream, even in events with two tracks there is only a 30\% probability that they are a signal pair.
    \item The number and momentum distribution of the tracks depends strongly on the D1 occupancy; if we plan to mix tracks from different events, we must make sure those events have similar occupancy.
\end{itemize}

\section{Combining Tracks to form a Spectrum}

As we have seen, an event may have 0, 1, 2, or more reconstructed tracks within it.  We can separate those tracks into two groups: {\em signal tracks} which have been produced by a muon from a $J/\psi$, $\psi'$, or Drell-Yan decay; and {\em background tracks} that have arisen from any other mechanism.  The distinguishing characteristic of signal tracks is that they come in pairs comprising one positive and one negative track. The background tracks do not share a physics vertex with any other track.
N.B.: an individual track from a $J/\psi$, $\psi'$, or Drell-Yan decay will be a background track if the other track in the pair was not reconstructed.

Within each event, we need to create all possible pairs of positive and negative tracks in search of the dimuon pairs from the signal sources.  In the process of doing so, we will combine signal tracks with background tracks, and also combine background tracks with other background tracks, thus forming the {\em combinatoric} background in the spectrum.  It is important to keep in mind that the combinatoric background contains a contribution from signal tracks as well as background tracks.

\section{Mimicking the Track-Pairing Process to Estimate the Combinatoric Background}
\label{sec:procedure}

To estimate the combinatoric background correctly, we need to combine tracks from the same populations of positive and negative tracks as found in the top/bottom data stream, but without the possibility of making any signal pairs.  We have seen that the wire chamber occupancy of an event strongly influences the number and momentum distributions of tracks, so we need to combine tracks from events with similar occupancy.  We also want to make sure that each track is only combined with tracks from one event, since that is what we do when forming a spectrum using the original data.

These considerations lead to the following algorithm for mixing tracks from different events.

\begin{enumerate}
    \item Choose a single normal data run, lasting about 1 hour.
    \item Select top/bottom-triggered events.
    \item Order the events according to occupancy (low to high, for example).
    \item Put all positive tracks from event $i$ and all negative tracks from event $(i+1)$ into a single new {\em mixed} event.  This is implemented for all events, including those with no reconstructed tracks.  There might be zero positive tracks in a given event, for example.  These ``mixed events'' are placed into a file structure called a ``mixed run''.
    \item Subsequent processing for mixed events occurs with identical conditions as for normal events. Especially, we make sure that the two tracks forming a dimuon from both normal and mixed events must satisfy the top/bottom trigger condition. This last requirement is important so that we preserve the bias of the top/bottom trigger in the mixed events.
\end{enumerate}
In the discussion that follows, a ``normal run'' will contain ``normal events'' that came from the original data stream, while a ``mixed run'' will contain ``mixed events'' created according to the algorithm described above.

\section{Normalization of the Estimated Combinatoric Background}
\label{sec:norm}
In this mixing method, we obtain the correct absolute normalization for the mixed distribution.  This statement depends on four conditions:  
(1) The density of events with signal pairs in the data stream is very small, as required in Ref.~\cite{Crochet:2001qd};
(2) Sorting the events according to D1 occupancy before mixing ensures the similarity of two events being mixed;
(3) A given track is only combined with tracks from within one event.  It is either combined with tracks within its own event (a normal run), or with tracks from one similar event (a mixed run);
(4) The normal events and the mixed events are subject to exactly the same cuts in the subsequent analysis.  In particular, all track pairs must satisfy the original top/bottom trigger condition.

We show that this method provides an estimate of the combinatoric background that is statistically consistent with the actual background.
A "run" is a set of data collected with a specific trigger with the experimental conditions unchanged; in SeaQuest, a run lasted about one hour.  A particular run may have $N_E$ events.  Each event $i$ has zero or more reconstructed tracks, which are broken into four groups:
\begin{itemize}
    \item $s_i^+$ is the number of positive tracks from a signal ($J/\psi$, $\psi'$, or Drell-Yan) = 0 or 1.
    \item $s_i^-$ is the number of negative tracks from a signal ($J/\psi$, $\psi'$, or Drell-Yan) = 0 or 1.
    \item $b_i^+$ is the number of positive tracks from a background = 0, 1, 2, ...
    \item $b_i^-$ is the number of negative tracks from a background = 0, 1, 2, ...
\end{itemize}
The signal tracks (positive and negative) come from a correlated source and only appear in pairs in the same event; we always have $s_i^+=s_i^-$.  The background tracks come from uncorrelated sources.  If only one of a pair of signal tracks is reconstructed, it falls into the background category.  Then the total number of unlike-sign track pairs, $N_P$, in a normal run is
$$ N_P = \sum_{i=1}^{N_E} \left(s_i^+s_i^- +  s_i^+b_i^- + b_i^+s_i^- + b_i^+b_i^-\right). $$
The first term in the sum is special, because $s_i^+$ and $s_i^-$ come from a correlated source.  The sum over this term is the total number of
signal dimuon pairs in this run, $N_S$.
$$ N_S = \sum_{i=1}^{N_E} s_i^+s_i^- $$
The other three terms generate the combinatoric background, $N_C$.
$$ N_C = \sum_{i=1}^{N_E} \left(s_i^+b_i^- + b_i^+s_i^- + b_i^+b_i^-\right) $$
At this point it is appropriate to sort the events into similar groups; in the case of SeaQuest, this means sorting them according to the D1 chamber occupancy, $\omega$, from low to high.  Then the sum can be broken down into sub-sums where all events have the same occupancy.  The number of events at a given occupancy $\omega$ is $N_\omega$.
$$ N_C = \sum_{\omega=0}^{\omega_{\rm max}} \sum_{i=1}^{N_\omega} \left(s_i^+b_i^- + b_i^+s_i^- + b_i^+b_i^-\right)   $$
The numbers $s_i^+$, $b_i^-$, and so on are all small integers (typically no larger than 7, see Fig.~\ref{fig:mult_D1}) drawn from a distribution depending on the occupancy.  On the other hand, $N_\omega$ will tend to be large, certainly a few hundreds or thousands for the most popular occupancies in a run.  The sum over events with the same occupancy will sample all possible values of $s_i^+b_i^-$ (e.g.) many times.  Therefore, we can replace the sum with averages:
$$ N_C = \sum_{\omega=0}^{\omega_{\rm max}} N_\omega \left(\left<s^+b^-\right>_\omega + \left<b^+s^-\right>_\omega + \left<b^+b^-\right>_\omega\right),$$
where $\left<s^+b^-\right>_\omega$ is the average value of the product $s_i^+b_i^-$ at the given occupancy $\omega$, etc.  Then the total number of pairs in the run is
$$ N_P = N_S + N_C.$$
Now consider the total number of unlike-sign track pairs, $N_P'$, in a {\em mixed run}, where we have combined the positive tracks from event $i$ with the negative tracks from event $i+1$ sourced from a normal run.  We have sorted the events by occupancy, so that adjacent events contain tracks drawn from the same distributions.
$$ N_P' = \sum_{i=1}^{N_E} \left(s_i^+s_{i+1}^- +  s_i^+b_{i+1}^- + b_i^+s_{i+1}^- + b_i^+b_{i+1}^-\right) $$
The sum over the first term $s_i^+s_{i+1}^-$ is non-zero but may be negligible; tracks from signals are rare and are only found in pairs in the same event.  The probability that adjacent events in our mixed run will both have signal tracks is very small.  We call this the ``adjacent signals'' term, $N_{\rm AS}$, and we ignore this term for the moment; this is the most important quantitative requirement for the proposed mixing method to work properly.  
$$    N_{\rm AS}=\sum_{i=1}^{N_E} s_i^+s_{i+1}^- \approx 0  $$
The remaining three terms may be treated in the same way as in the normal run.
$$ N_C' = \sum_{i=1}^{N_E} \left(s_i^+b_{i+1}^- + b_i^+s_{i+1}^- + b_i^+b_{i+1}^-\right) = 
\sum_{\omega=0}^{\omega_{\rm max}} N_\omega \left(\left<s^+b^-\right>_\omega + \left<b^+s^-\right>_\omega + \left<b^+b^-\right>_\omega\right)$$
The sums $N_C$ (from the normal run) and $N_C'$ (from the mixed run) are equal in the limit of large statistics.  In the case of limited statistics, they will be equal within statistical uncertainties. Then to estimate the number of signal pairs, we need only subtract the mixed run from the normal run.
$$  N_P - N_P' = (N_S + N_C) - N_C' \approx N_S  $$
as $N_C \approx N_C'$ within uncertainties.

The question of whether the adjacent signals term $N_{\rm AS}$ can be ignored depends on two considerations: the probability $f$ that a given event will contain a signal pair, and the overall statistical significance of the experiment.  A numerical example is useful.  Suppose $f=0.01$ (1\% of events have signal pairs) and in total there are $10^4$ events.  The number of signal pairs is approximately 100.  The probability that two adjacent events have signal pairs is $f^2$, so the number of adjacent signal pairs is just 1.  Compared to the statistical uncertainty in the number of signals (10), the adjacent signals term may be ignored.  On the other hand, if there are $10^6$ events, then the number of signal pairs is $10^4$ and the number of adjacent signal pairs is 100, which is the same order as the uncertainty in the number of signal pairs.  In this second case the adjacent signals term needs to be taken into account.  We will show that the effect of the adjacent signals can be quantified via simulation and embedding, and so may be corrected for.

\section{Simple Models Illustrating the Proposed Event-Mixing Method}
\label{sec:simple_model}

To demonstrate the abilities and limitations of this mixing method, we create simple statistical models of the event stream.  We will use the simple shape $\exp(-An)$ for the parent distribution for signals and backgrounds; $n$ is the number of signal pairs or background tracks, and $A$ is the shape constant of the curve; increasing the value of $A$ means the signal or background becomes more rare.

In our first simple model, we have chosen the values of $A$ so that the signal pairs are very rare compared to the background tracks. 
\begin{itemize}
    \item[] $P_S \propto \exp(-7n_s)$ -- probability of the number of signal pairs in the event
    \item[] $P_P \propto \exp(-4n_p)$ -- probability of the number of positive background tracks in the event
    \item[] $P_N \propto \exp(-5n_n)$ -- probability of the number of negative background tracks in the event
\end{itemize}
In each event, integer values of $n_s$, $n_p$, and $n_n$ are chosen based on these parent probability distributions, determining the number of signal and background tracks in the event. For example, a selection of $(n_s,n_p,n_n)=(1,1,2)$ would mean $(s^+, s^-, b^+, b^-)=(1,1,1,2)$.  Then the following process is followed.
\begin{enumerate}
    \item Create a list of 100,000 events using these probability distributions.
    \item Calculate the total number of track pairs, the true number of signal pairs, the true number of combinatoric pairs, the estimated number of combinatoric pairs using the proposed method, and the difference in the number of combinatoric pairs (true - estimated).
    \item Repeat steps 1 and 2 10,000 times, and histogram the above quantities.
\end{enumerate}

\begin{figure}[ht]
    \centering
    \includegraphics[width=0.45\linewidth]{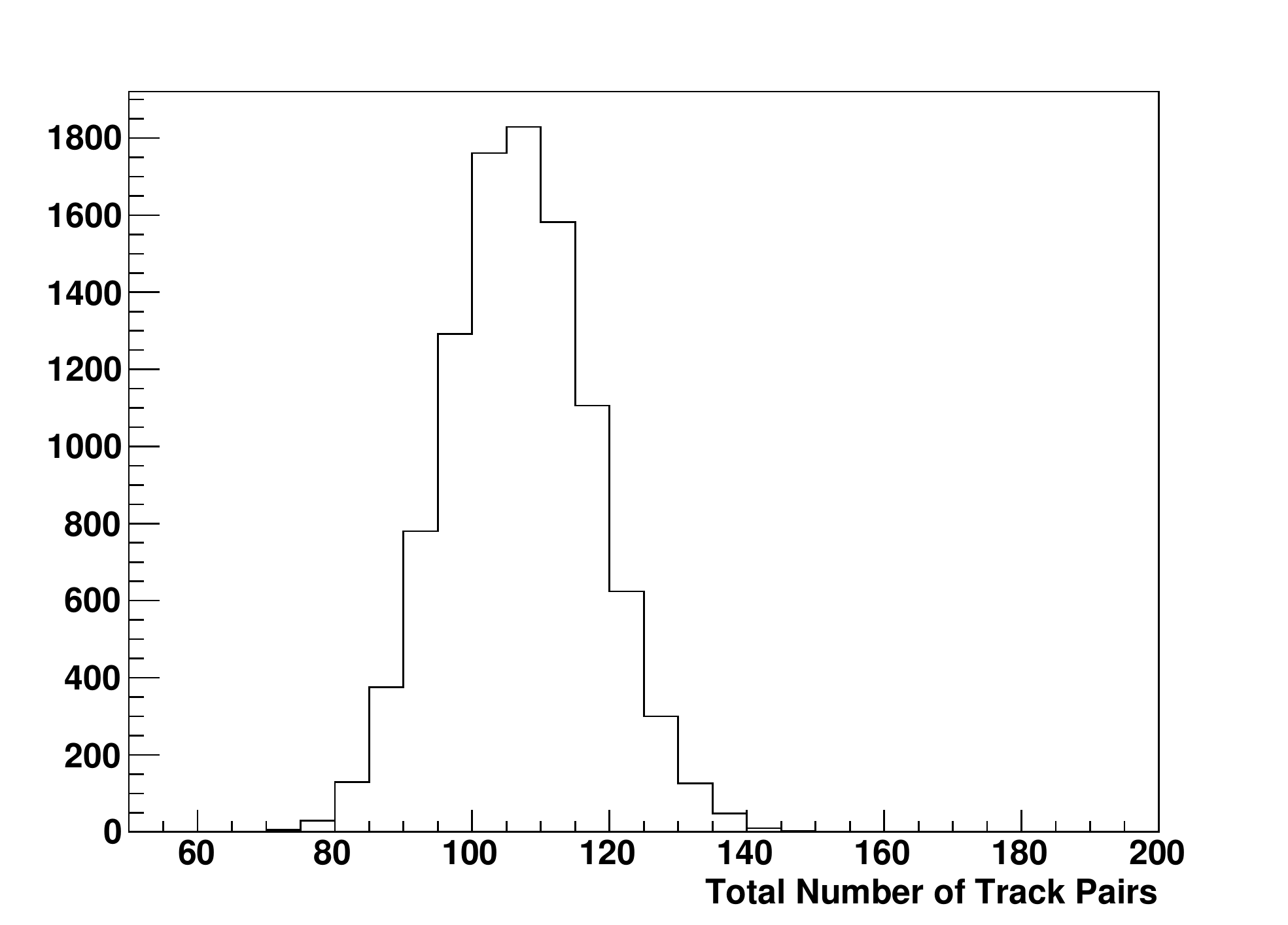}
    \includegraphics[width=0.45\linewidth]{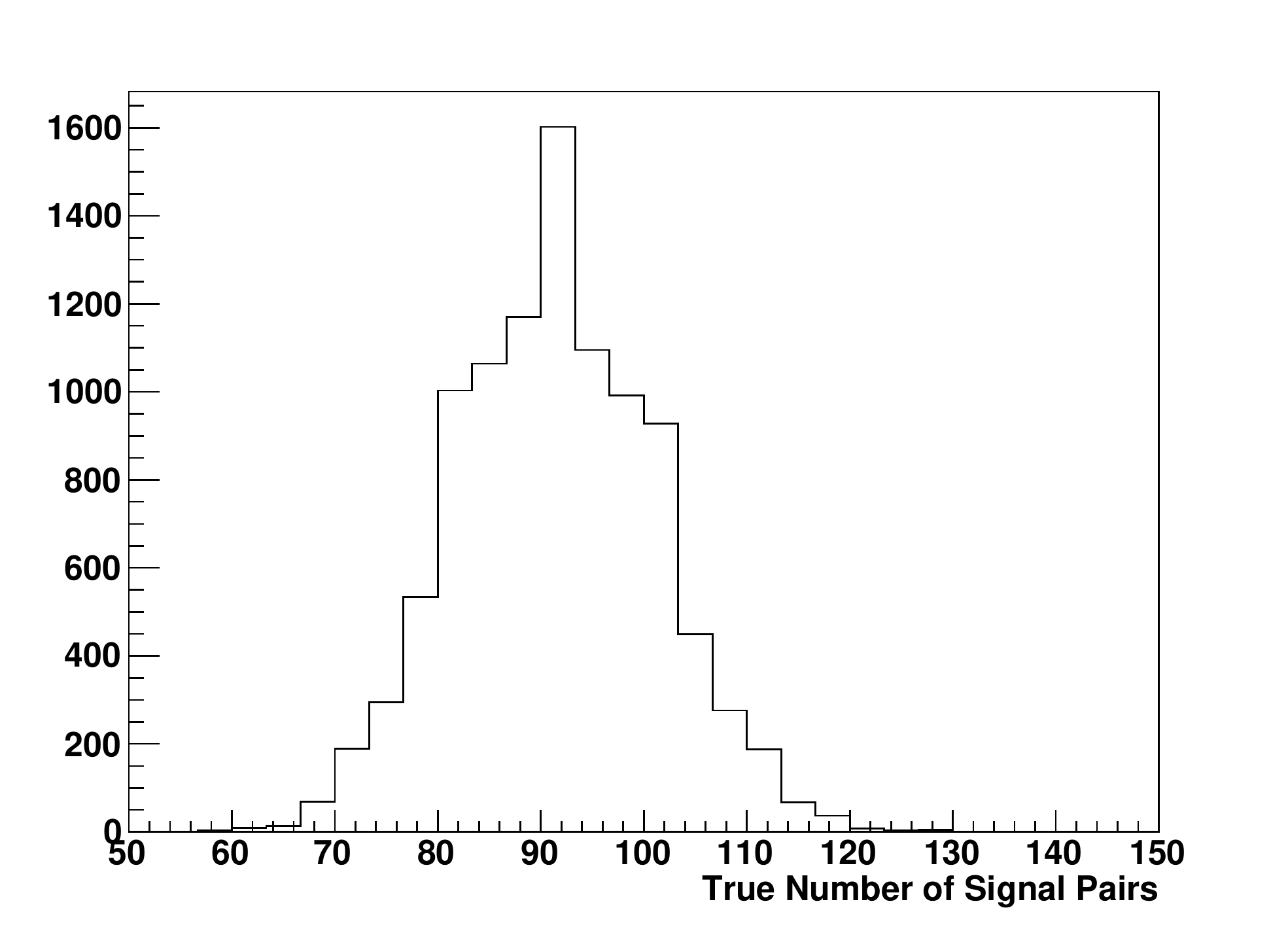}\\
    \includegraphics[width=0.45\linewidth]{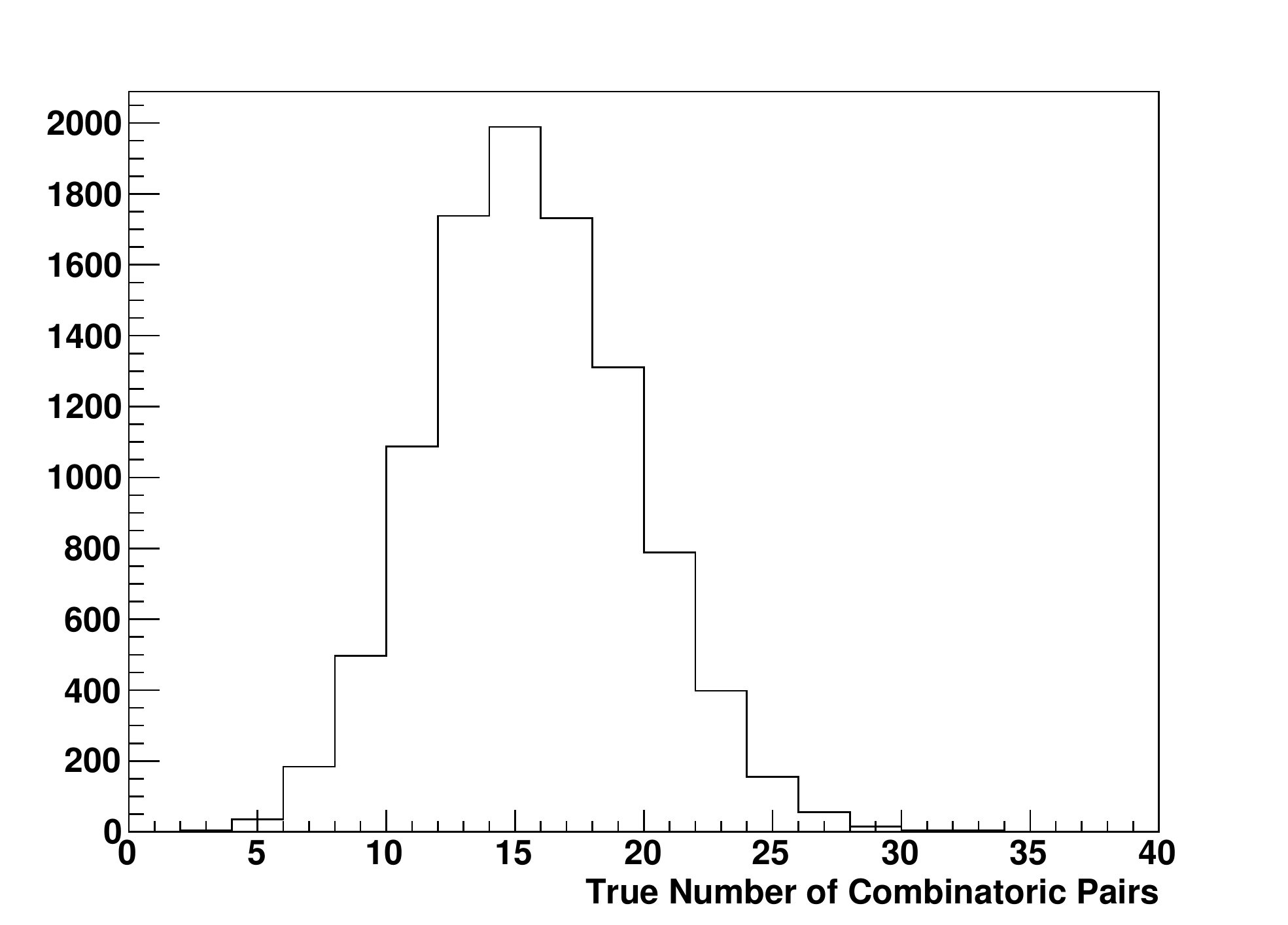}
    \includegraphics[width=0.45\linewidth]{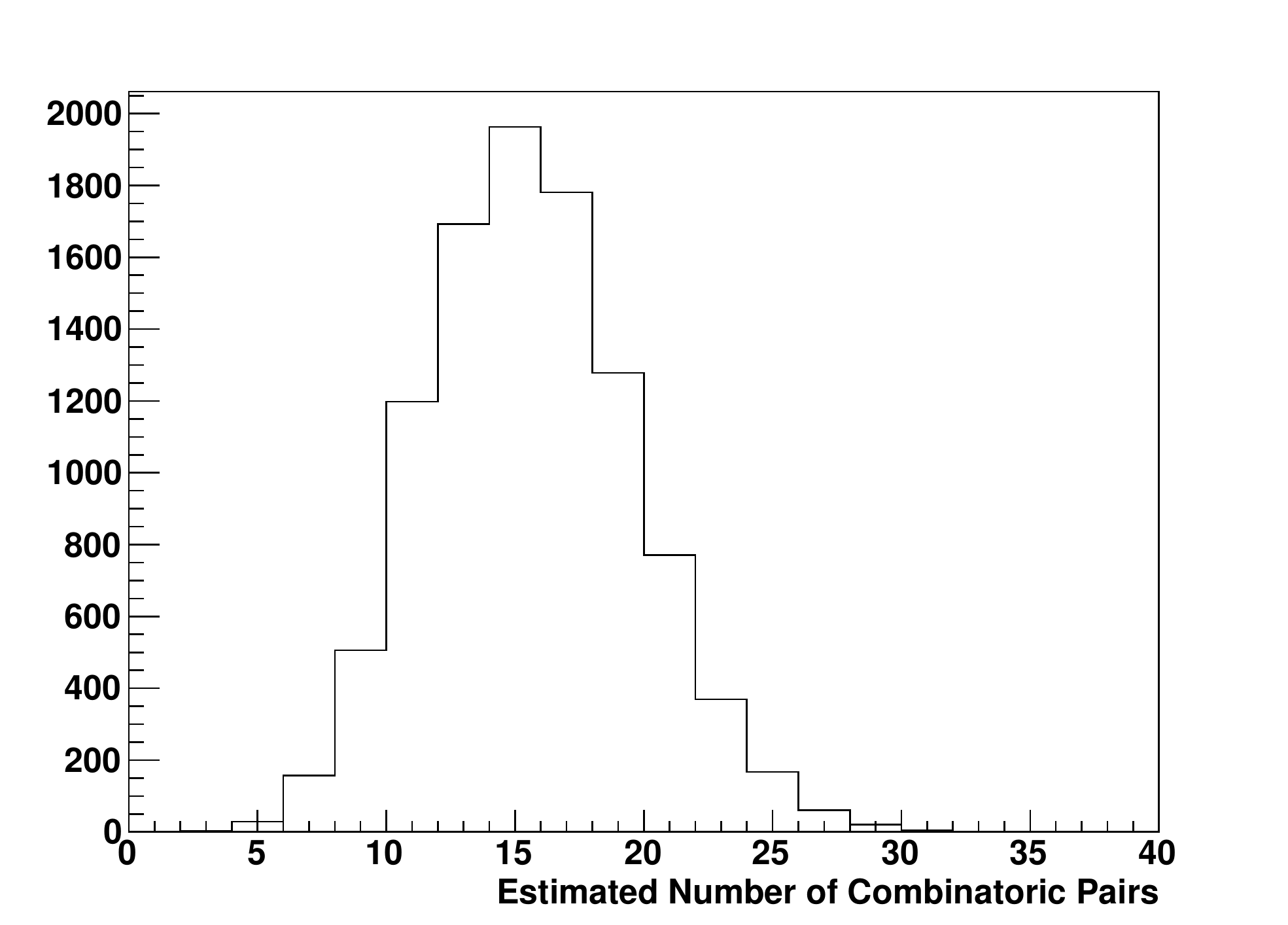}\\
    \includegraphics[width=0.45\linewidth]{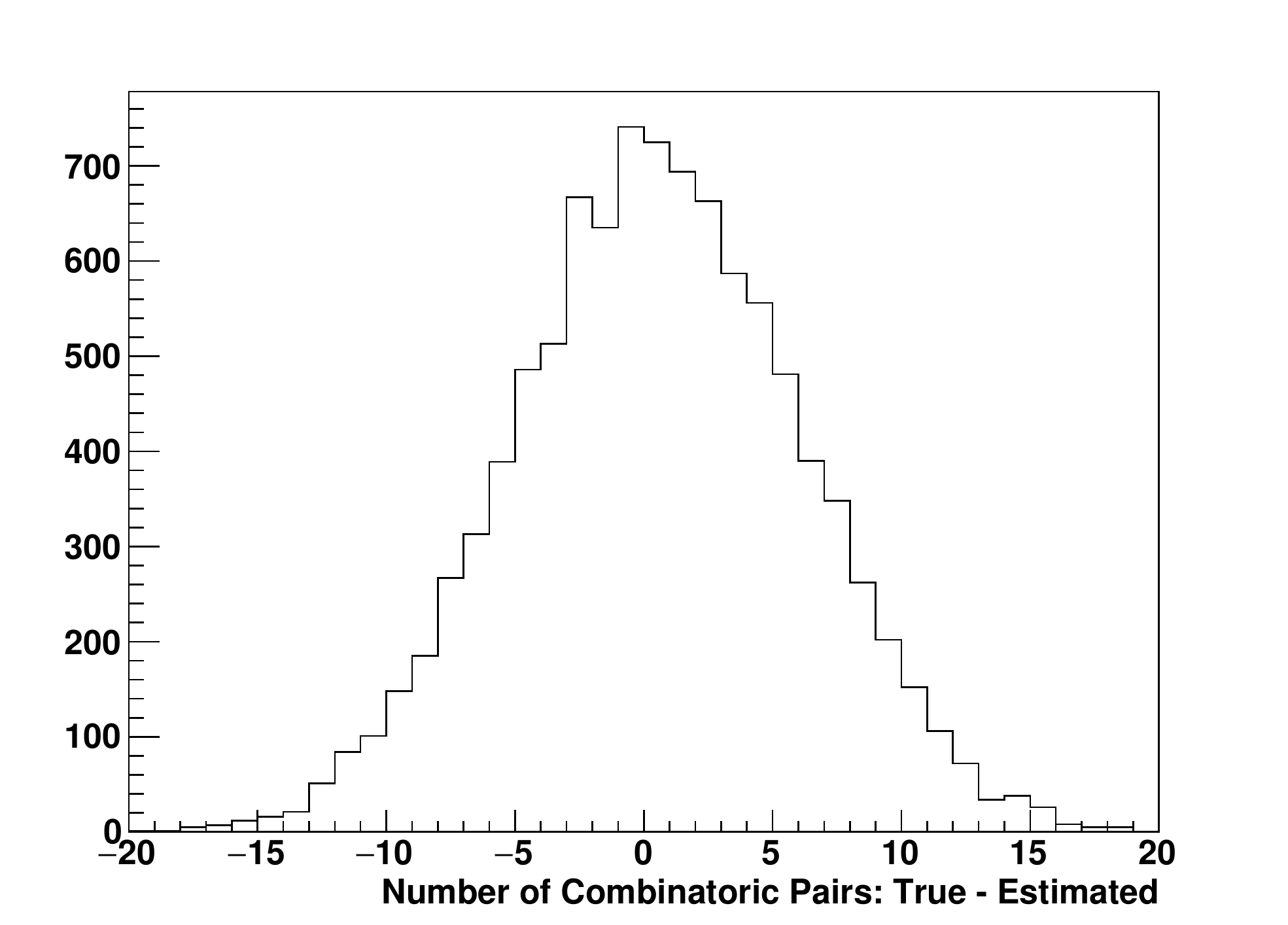}
    \caption{Results from first simple model.  Top Left:  Histogram of the total number of track pairs.  Top Right:  True number of signal pairs.  Middle Left:  True number of combinatoric background pairs.  Middle Right:  Estimated number of combinatoric background pairs.  Bottom:  Difference between true and estimated number of combinatoric background pairs.}
    \label{fig:sim1}
\end{figure}

The results are illustrated in the histograms in Fig.~\ref{fig:sim1}.  The average number of track pairs per 100,000 events is 105, of which on average there are 91 true signal pairs.  The histograms for the true and estimated numbers of combinatoric background pairs are extremely similar, with the same average of 15.  The difference between true and estimated combinatoric background pairs is centered about zero, with a root-mean-square deviation of 5.6 consistent with the difference of the averages; the statistical error on $15-15$ would be $\sqrt{15+15}=5.5$.  We see that the proposed method works very well in this scenario.

\begin{figure}[ht]
    \centering
    \includegraphics[width=0.45\linewidth]{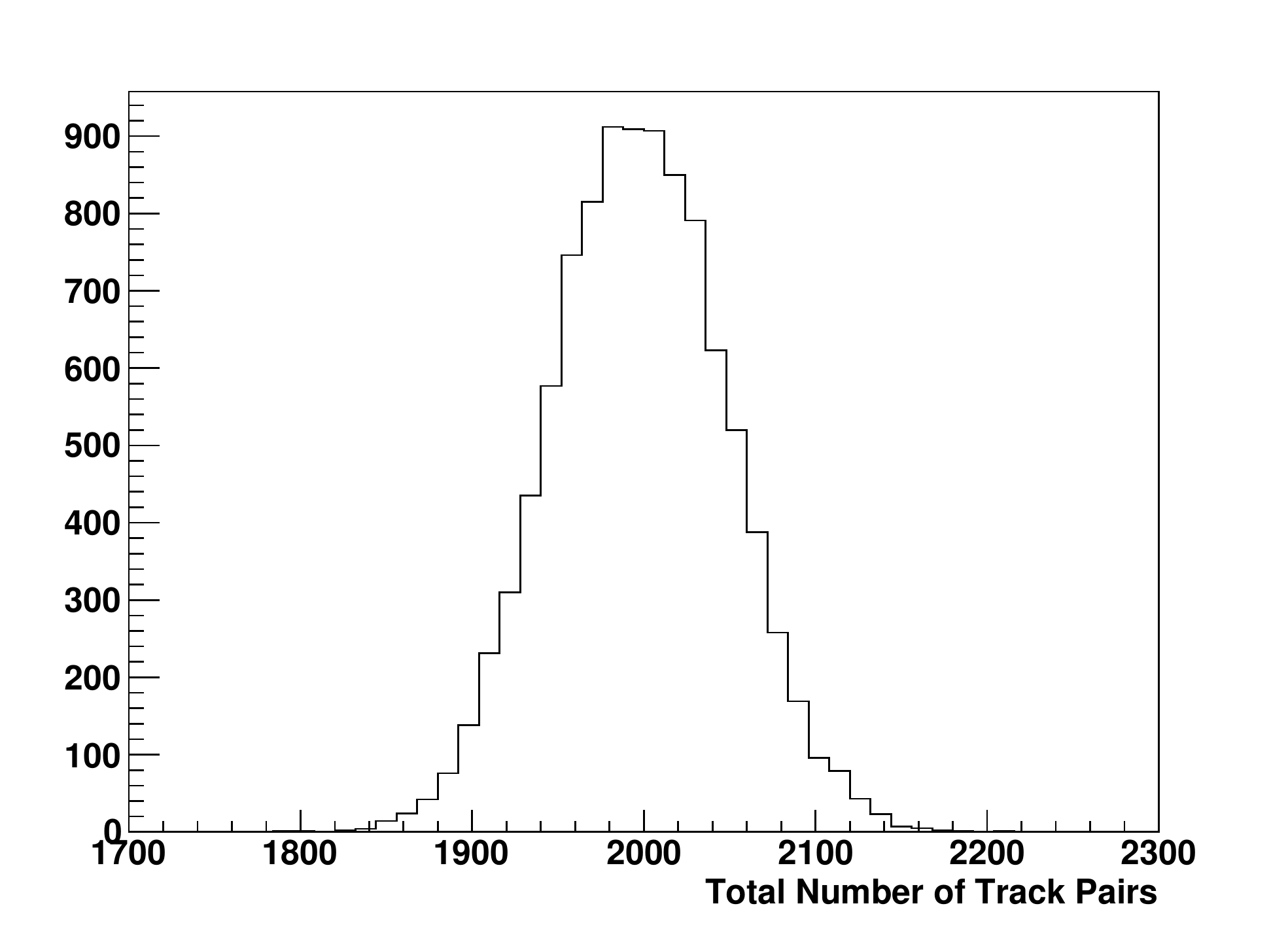}
    \includegraphics[width=0.45\linewidth]{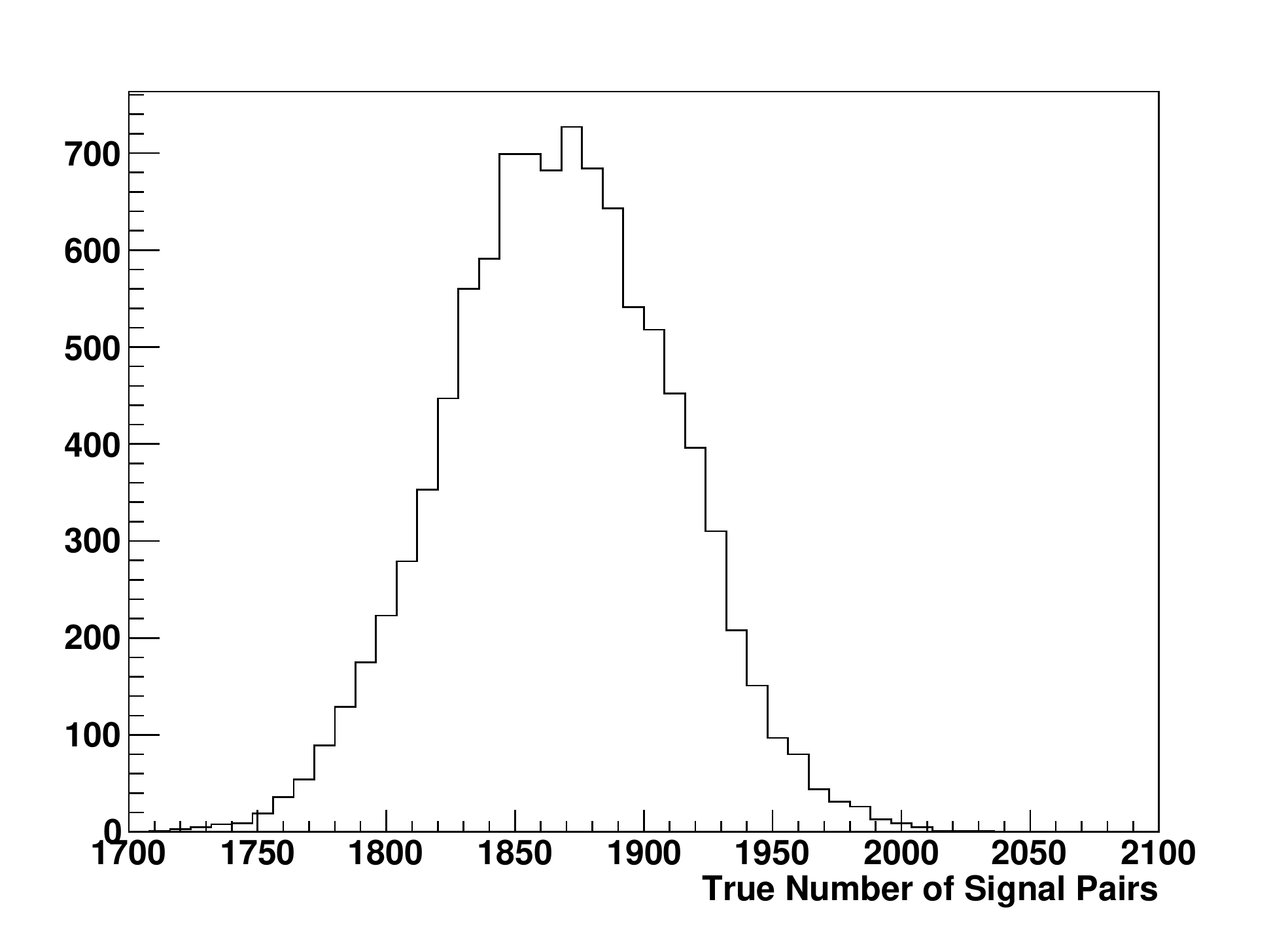}\\
    \includegraphics[width=0.45\linewidth]{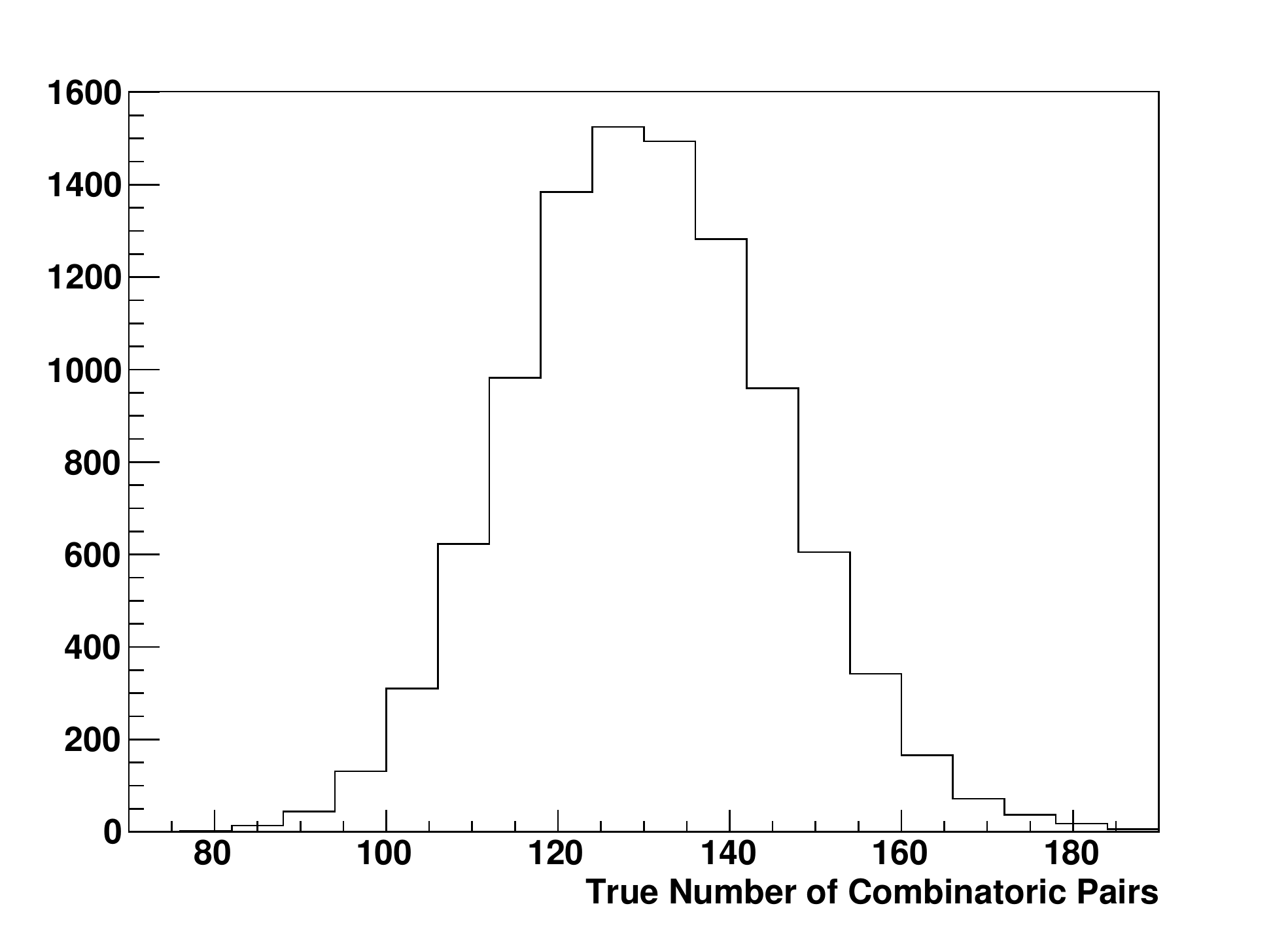}
    \includegraphics[width=0.45\linewidth]{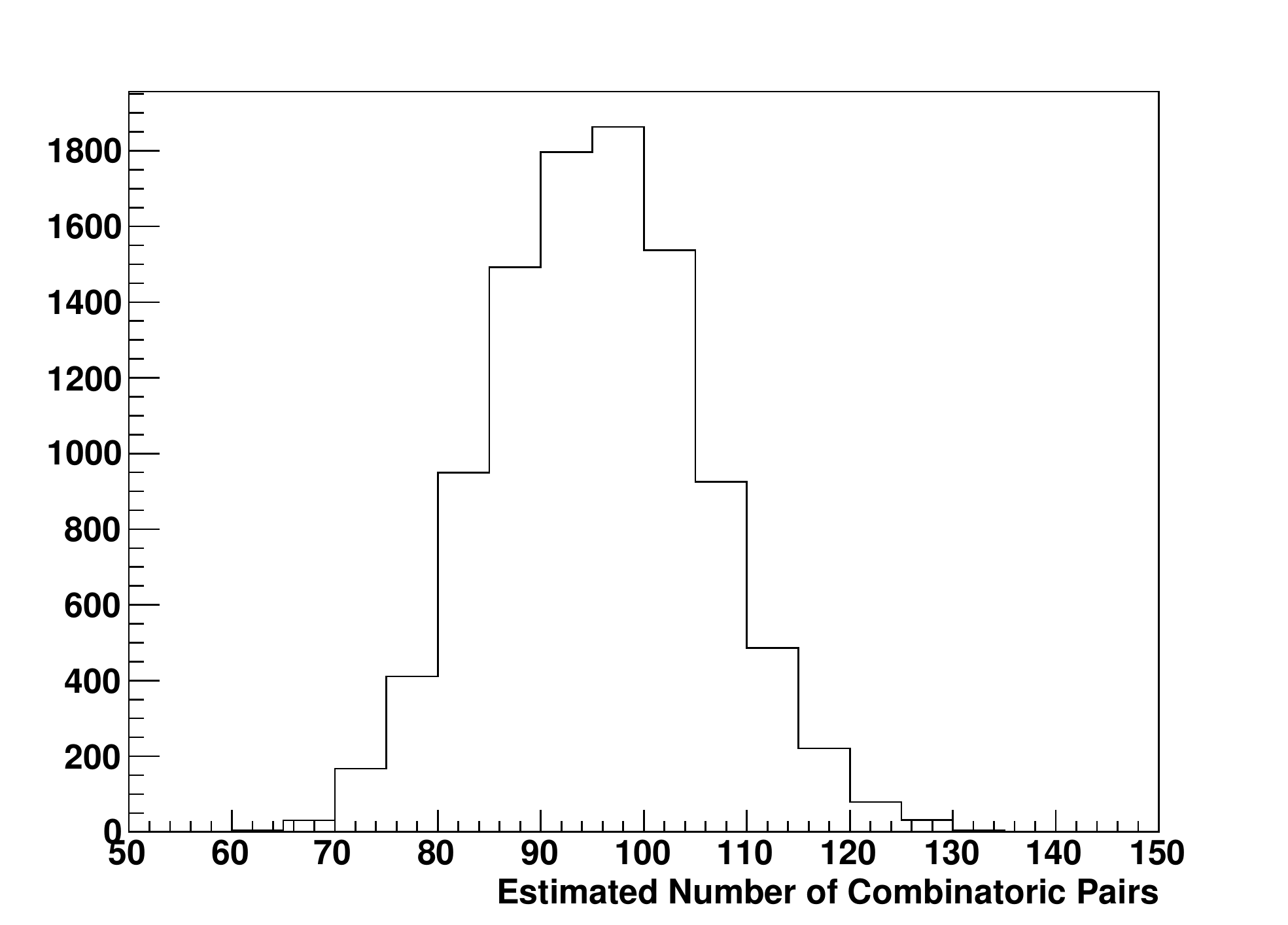}\\
    \includegraphics[width=0.45\linewidth]{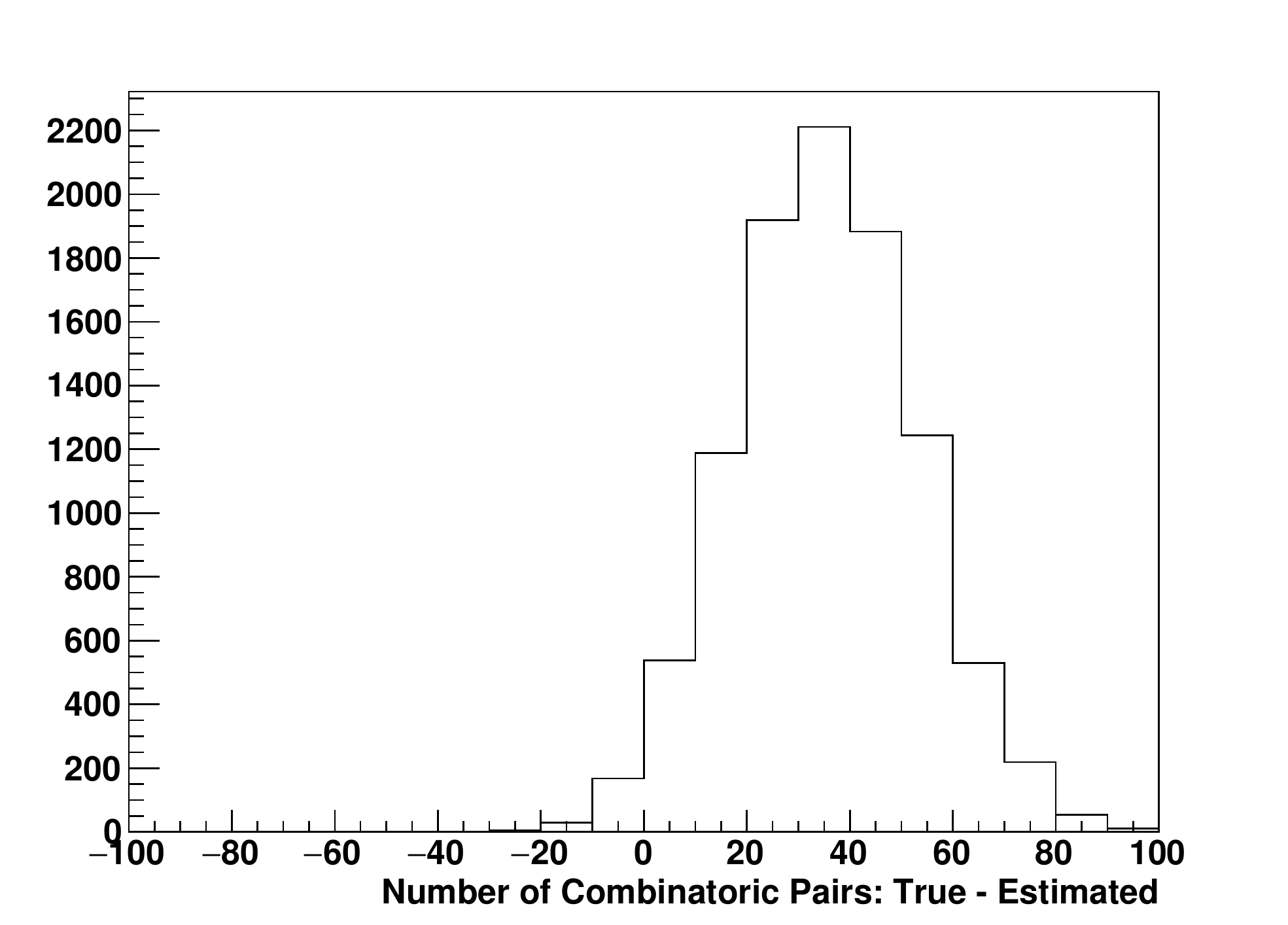}
    \caption{Results from second simple model.  Top Left:  Histogram of the total number of track pairs.  Top Right:  True number of signal pairs.  Middle Left:  True number of combinatoric background pairs.  Middle Right:  Estimated number of combinatoric background pairs.  Bottom:  Difference between true and estimated number of combinatoric background pairs.}
    \label{fig:sim2}
\end{figure}

In our second simple model, we increase the rate of signal pairs so that it is comparable to the rate of background tracks; this violates one of the assumptions of the proposed mixing method and so we expect this to fail.  The parent distributions are now:
\begin{itemize}
    \item[] $P_S \propto \exp(-4n_s)$ --- compare to $\exp(-7n_s)$ in the first model
    \item[] $P_P \propto \exp(-4n_p)$ --- same as in first model
    \item[] $P_N \propto \exp(-5n_n)$ --- same as in first model
\end{itemize}

The same procedure is followed, and the results are shown in Fig.~\ref{fig:sim2}.  With a much greater density of signal pairs, the average total number of track pairs is increased to approximately 2,000, of which on average there are approximately 1,870 signal pairs.  The number of true combinatoric pairs therefore averages around 130, but the proposed method {\em underestimates} this to be about 95.  The reason for the underestimation is we now can have more than one signal pair per event, and the proposed method does not reproduce the extra combinatoric background created by the signal pairs among themselves.  The method fails if the signal-to-background ratio is too high.

\begin{figure}[ht]
    \centering
    \includegraphics[width=0.45\linewidth]{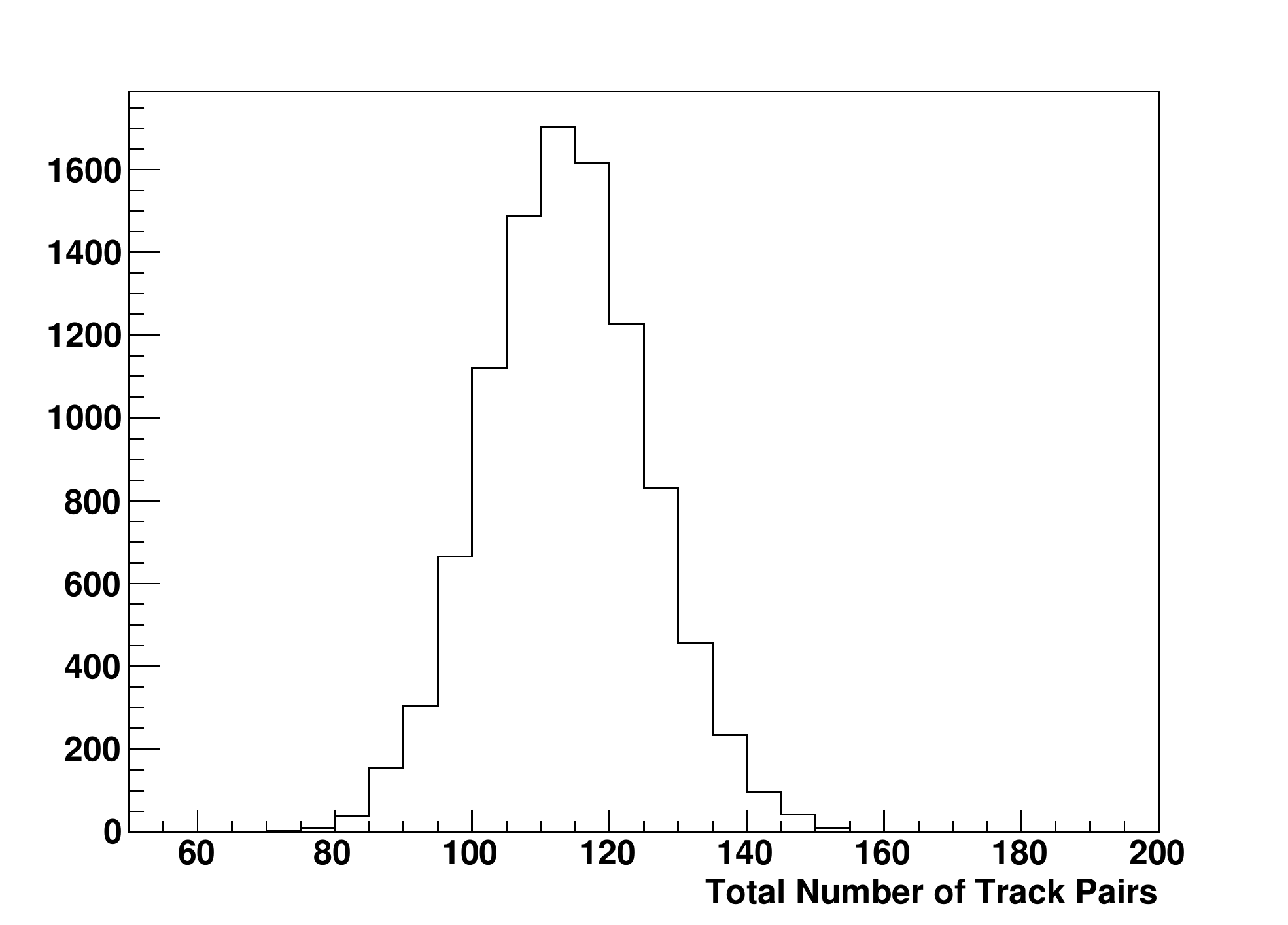}
    \includegraphics[width=0.45\linewidth]{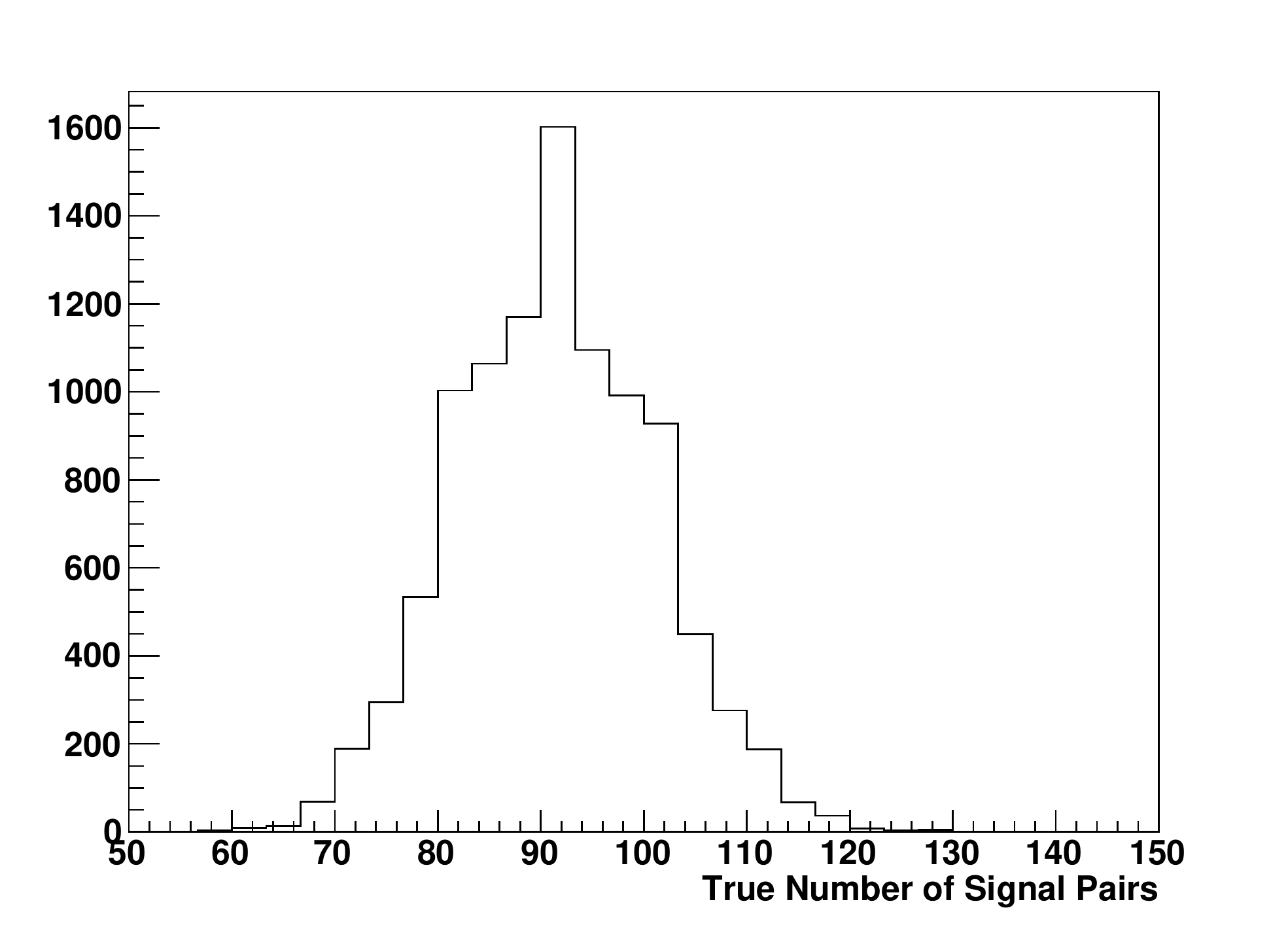}\\
    \includegraphics[width=0.45\linewidth]{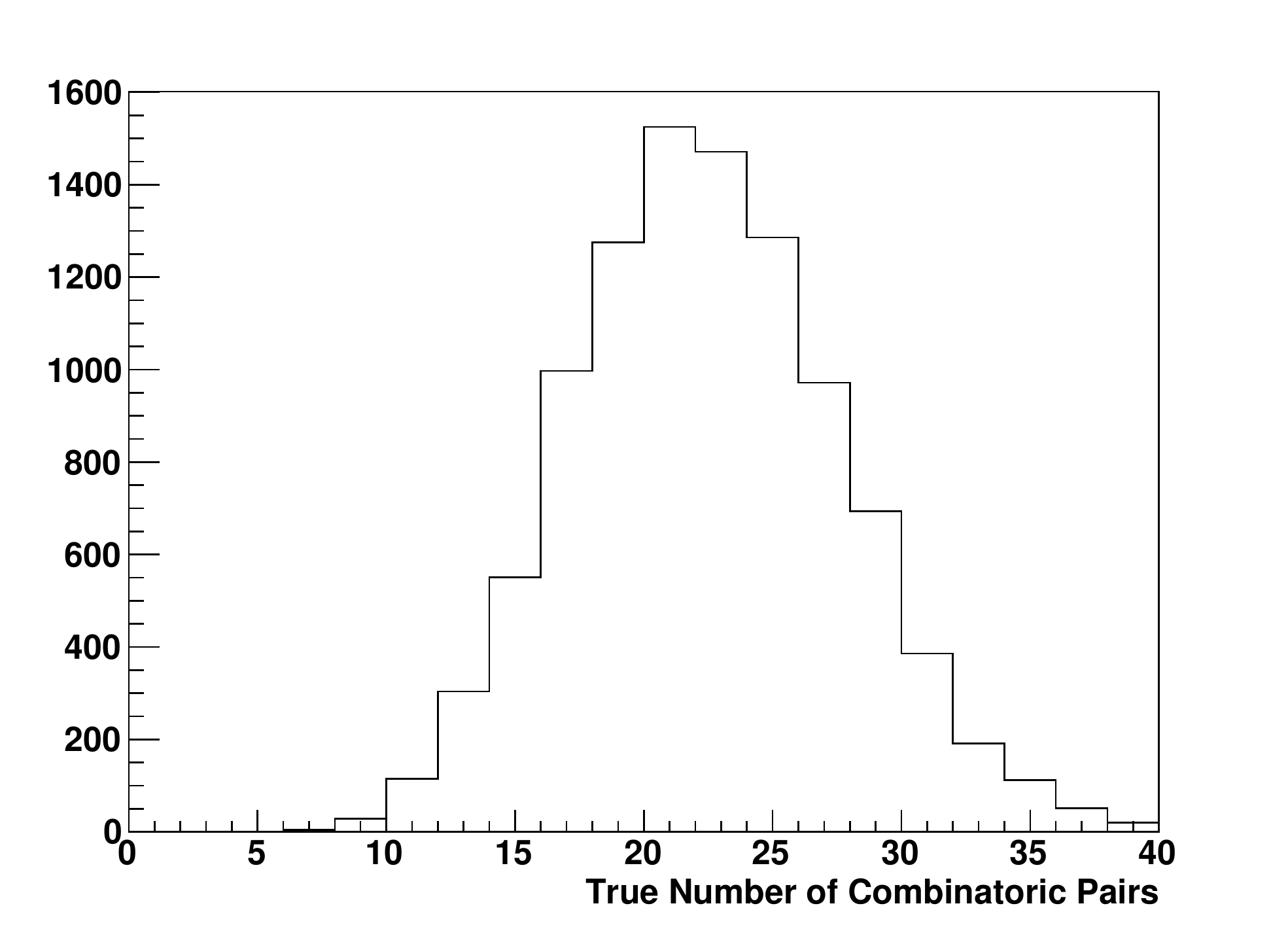}
    \includegraphics[width=0.45\linewidth]{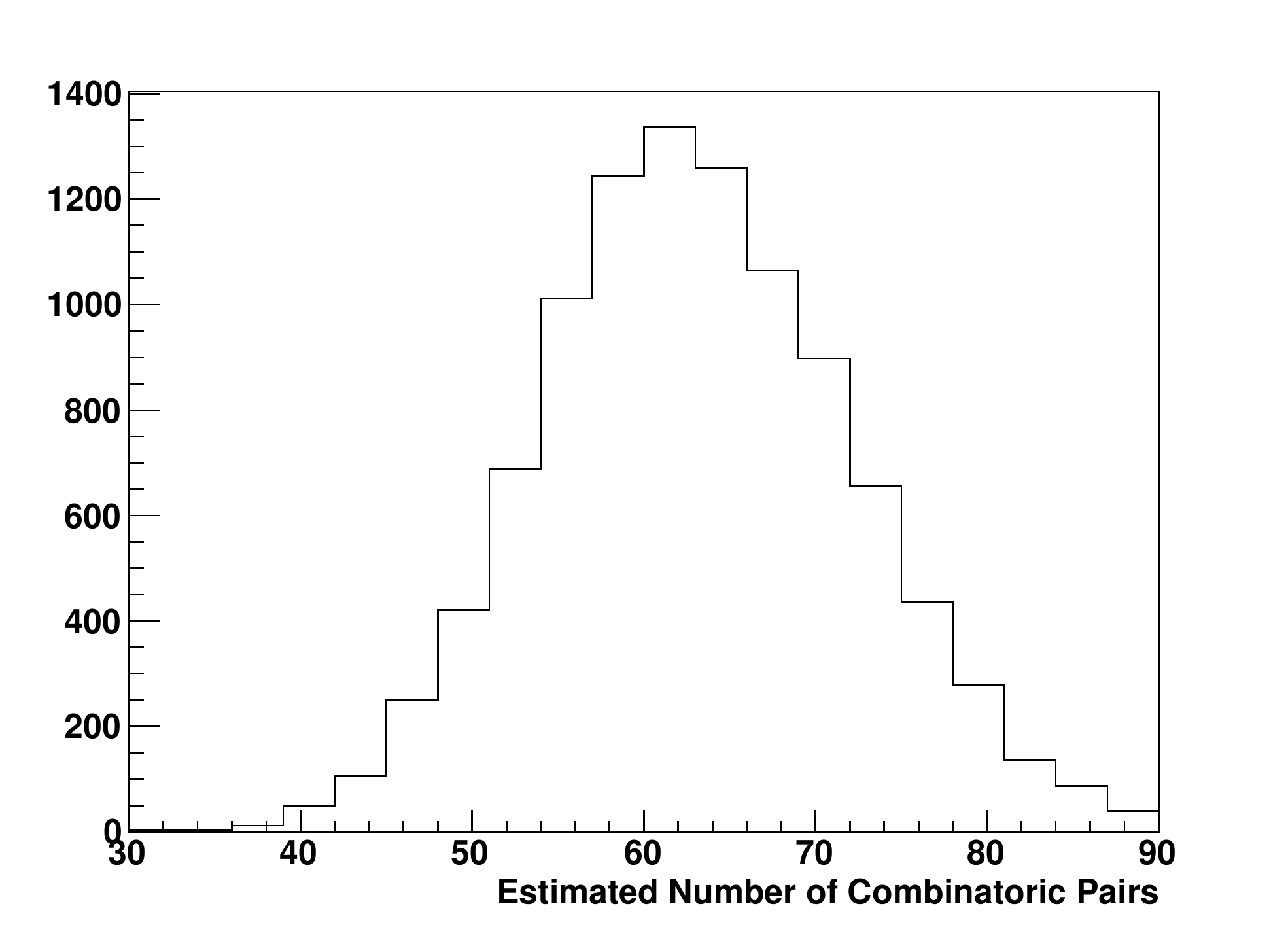}\\
    \includegraphics[width=0.45\linewidth]{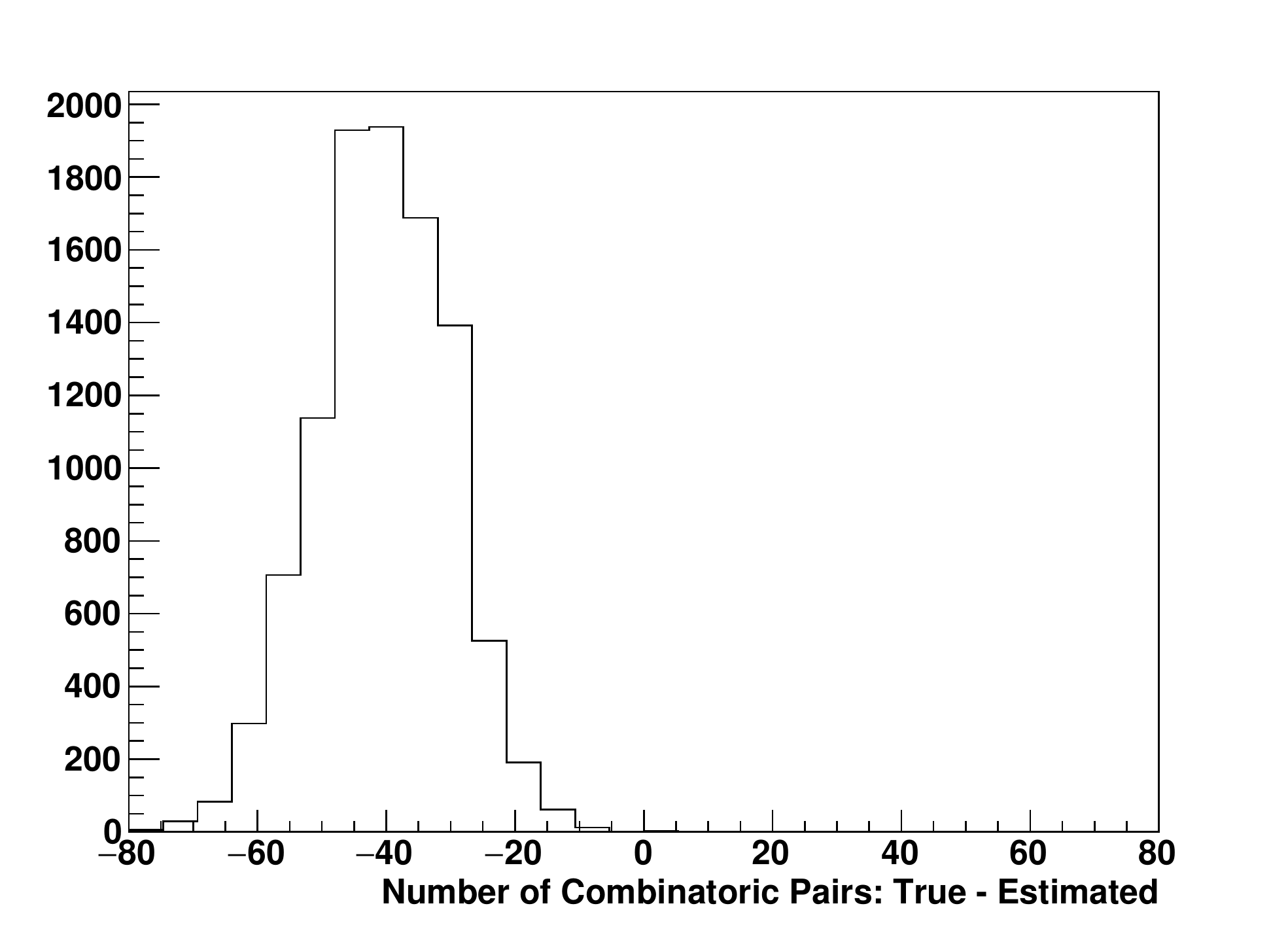}
    \caption{Results from third simple model.  Top Left:  Histogram of the total number of track pairs.  Top Right:  True number of signal pairs.  Middle Left:  True number of combinatoric background pairs.  Middle Right:  Estimated number of combinatoric background pairs.  Bottom:  Difference between true and estimated number of combinatoric background pairs.}
    \label{fig:sim3}
\end{figure}

In our third simple model, we make the signal pairs rare again, but we make the background track distributions alternate between even-numbered and odd-numbered events.
\begin{eqnarray*}
P_S & \propto \exp(-7n_s) & \text{in all events}\\
P_P & \propto \exp(-4n_p) & \text{in even-numbered events}\\
& \propto \exp(-6n_p) & \text{in odd-numbered events}\\
P_N & \propto \exp(-5n_n) & \text{in even-numbered events}\\
& \propto \exp(-4n_n) & \text{in odd-numbered events}
\end{eqnarray*}
In this scenario, when we mix tracks from adjacent events in the list of events, we will be mixing tracks from events with different track distributions, which violates one of the assumptions of the proposed mixing method, and so we expect this to fail.

In Fig.~\ref{fig:sim3}, we see that the proposed method vastly overestimates the actual number of combinatoric tracks, because of the mis-match in track distributions that occurs during the mixing.  The proposed method fails if you mix tracks from different distributions.

\section{Tests of the Event-Mixing Method Using SeaQuest Data}
\label{sec:cons_check}
We performed various tests, using actual SeaQuest track data, to check the validity of the this mixing method. There were two distinct types of tests.  Both types employ simulated track data embedded into the actual data stream.
\begin{itemize}
    \item[] Type 1 Test
\begin{itemize}
    \item We start with a {\em mixed run} as described above.  This is a set of events containing tracks from a real run, sorted by occupancy, and then the positive tracks from one event mixed with the negative tracks from the next event. There is no physics signal in a mixed run; all possible unlike-sign track pairs within mixed events are uncorrelated pairs. 
    \item Into these events, we embed reconstructed track pairs from a simulated signal. (The simulated signal is called ``GMC'' for ``generated Monte Carlo.'') A track pair is embedded into every $n^{\rm th}$ event; the number $n$ is chosen so that the embedded signal is sparse (on order of a few percent) in the same way that real signal pairs are sparse (about 5\%) in the real data. 
    \item We analyze this set of events containing embedded tracks like it was a normal run.  First, we loop over the events and form unlike-sign track pairs to make a spectrum; then we follow our procedure to mix tracks from different events, and produce a combinatoric background spectrum; we subtract the second spectrum from the first; the result should be the signal that we embedded.
\end{itemize}
That type of test was done with two different simulated signals; in one case with a broad set of generated Drell-Yan events, and in a second case with a ``resonance'' at an invariant mass of 6 GeV/$c^2$.
\item[] Type 2 Test
\begin{itemize}
\item We start with a {\em normal} run, as described above.  This is a set of events containing tracks from a real run, containing physics signal tracks as well as background tracks. 
\item Into these events, we embed reconstructed track pairs from a simulated (GMC) signal. A track pair is embedded into every $n^{\rm th}$ event, in the same manner as in the Type 1 Test. 
\item Now we perform two analyses.
\begin{enumerate}
    \item One is done with the original normal run (without embedded tracks), using the event mixing method.  We calculate a spectrum from unlike-sign pairs (Yield1) and also the corresponding combinatoric background (Comb1) and then subtract the combinatoric background: Signal1 = Yield1 -- Comb1.
    \item The second analysis is done with the normal run containing the embedded GMC tracks.  We calculate a spectrum from unlike-sign pairs (Yield2) and also the corresponding combinatoric background (Comb2) and then subtract the combinatoric background: Signal2 = Yield2 -- Comb2.
\end{enumerate}
\item Now we subtract Signal1 from Signal2 and the result should be the embedded GMC signal.  
\end{itemize}
This more elaborate test was performed with three different simulated signals: a resonance at 3.14 GeV/$c^2$, a resonance at 6 GeV/$c^2$, and a broad spectrum of Drell-Yan events.
 
\end{itemize}

An important feature available to us in these embedding schemes is that we can turn off the adjacent signals term $N_{\rm AS}$ by embedding the simulated tracks pairs at fixed intervals in the sequence of events; this means there are certainly no adjacent events with embedded signals.  Alternatively, we can embed randomly in the event stream, and thus turn on the adjacent signals term.  Looking at both kinds of embedding (fixed vs. random) enables us to quantify the effect of adjacent signals.

The results of one of these tests are given in the following subsection.  The balance of the test results are given in the Appendix.

\subsection{Type 2 Test with 6 GeV/$c^2$ Resonance Embedded Signal}
\label{subsec:6GeV}

Here we explain in detail a Type 2 test with a simulated 6 GeV/$c^2$ resonance signal. In this test, the embedded signals are placed at fixed intervals in the event stream; this means the effect of the adjacent signals term $N_{\rm AS}$ is turned off, and we are demonstrating that the $N_C'$ term generated in the mixed data correctly estimates the combinatoric background $N_C$ generated from the normal data.  We used 10 normal runs of SeaQuest data, using top/bottom-triggered events. The reconstructed tracks of the simulated signal were obtained by running generated  dimuon pairs with an invariant mass of 6 GeV/$c^2$ through the full detector simulation and reconstruction. For this analysis, no further cuts and conditions are applied other than requiring the dimuons to satisfy the top/bottom trigger and to form a proper dimuon vertex. The following are the details for this test.

\begin{figure}[ht]
\centering
\begin{subfigure}{.5\textwidth}
  \centering
  \includegraphics[width=1.\linewidth]{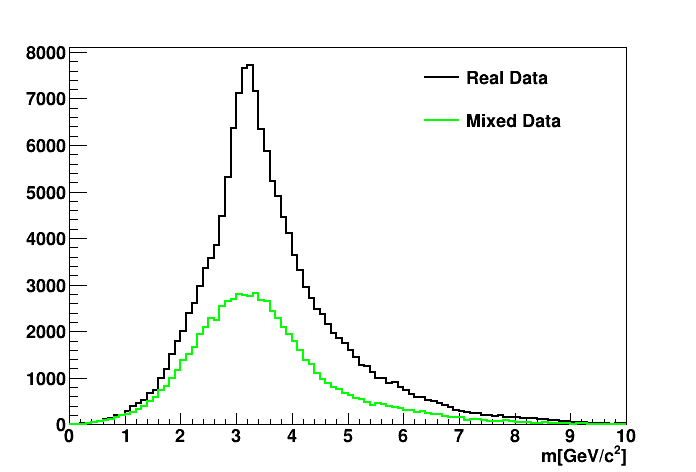}
\end{subfigure}%
\begin{subfigure}{.5\textwidth}
  \centering
  \includegraphics[width=1.\linewidth]{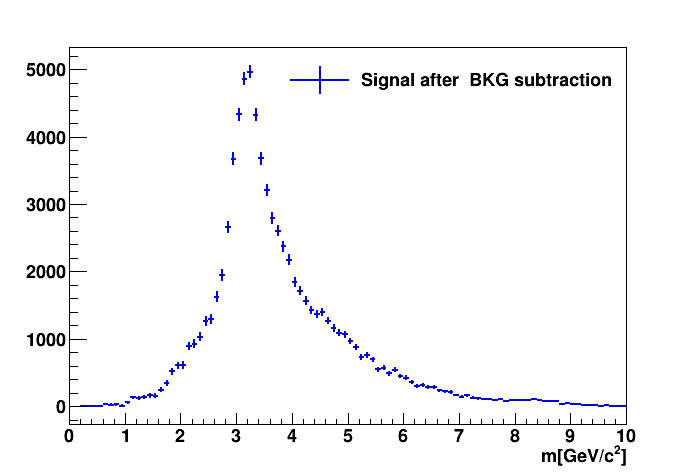}
\end{subfigure}
\caption{Left: Invariant mass distribution from the 10 runs of normal data (black) and corresponding mixed (green) distribution. Right: Signal obtained by subtracting mixed distribution (BKG) from data.  Blue = Black -- Green.}
\label{fig:rs67-data}
\end{figure}

\begin{figure}[ht]
\centering
\includegraphics[width=0.5\linewidth]{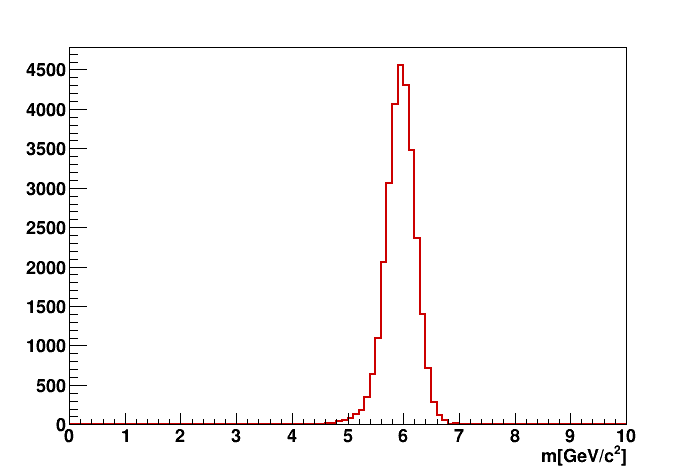}
\caption{ The simulated 6-GeV/$c^2$ resonance signal that will be embedded into the data displayed in Fig.~\ref{fig:rs67-data}.}
\label{fig:embed_signal}
\end{figure}

\begin{figure}[ht]
\centering
\begin{subfigure}{.5\textwidth}
  \centering
  \includegraphics[width=1.\linewidth]{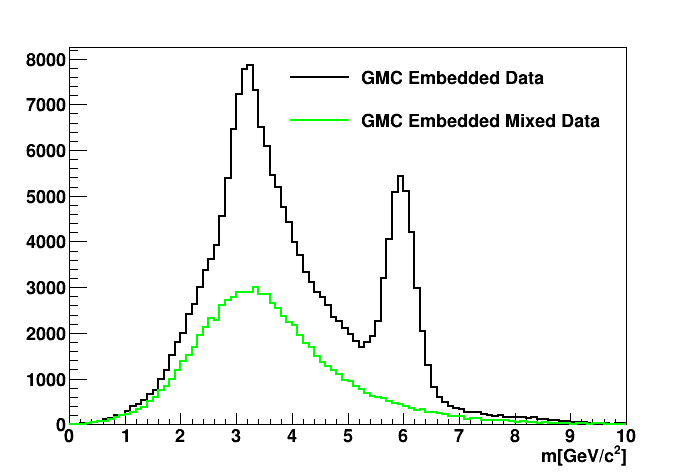}
\end{subfigure}%
\begin{subfigure}{.5\textwidth}
  \centering
  \includegraphics[width=1.\linewidth]{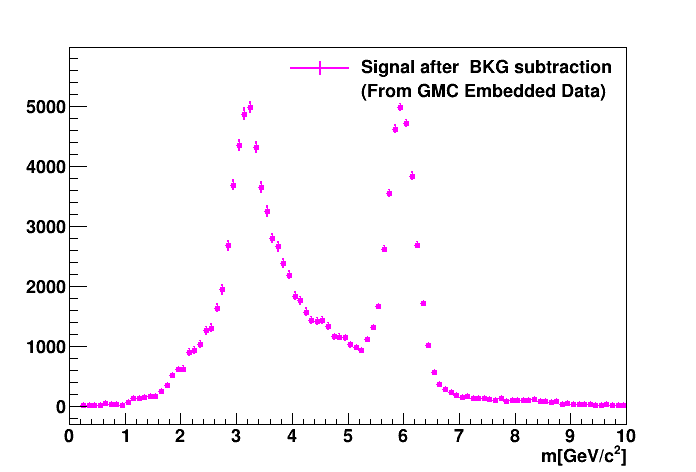}
\end{subfigure}
\caption{Left: Dimuon distribution (black) after the simulated signal from Fig.~\ref{fig:embed_signal} is embedded into the data of Fig.~\ref{fig:rs67-data}, and the corresponding mixed distribution (green). Right: Signal obtained by subtracting mixed distribution from data.  Magenta = Black -- Green.}
\label{fig:embedded-data}
\end{figure}

\begin{itemize}
    \item We took 10 normal data runs. 
    \item The mixing method was applied to each individual run, and then the results combined together. The left side of Fig.~\ref{fig:rs67-data} shows the total dimuon mass distribution for real (black line, Yield1) and mixed events (green line, Comb1) from the data.
    \item The total signal (Signal1) is obtained by subtracting the mixed distribution (Comb1) from the real data distribution (Yield1); this is shown in the right side of Fig.~\ref{fig:rs67-data}.  This spectrum contains a strong $J/\psi$ peak near 3 GeV/$c^2$, a shoulder (barely visible) from $\psi'$ production just above 3 GeV/$c^2$, and a continuum of Drell-Yan events.  The spectrum falls to zero beyond 1 GeV/$c^2$ and 9 GeV/$c^2$ due to the acceptance of the spectrometer.
    \item The reconstructed simulated dimuon signal with mass of 6 GeV/$c^2$ is shown in Fig.~\ref{fig:embed_signal}.  The tracks for the simulated signal are embedded into every $50^{\rm th}$ event in the 10 runs.
    \item The mixing method is applied to individual runs containing the embedded tracks. The total dimuon distribution (Yield2) after embedding as mentioned above is shown in the black line in the left side of Fig.~\ref{fig:embedded-data}. In the same figure, the green line shows the mixed distribution (Comb2) after applying the mixing procedure.
    \item The total signal (Signal2) obtained from subtracting the mixed distribution from the embedded data distribution is shown in the right side of Fig.~\ref{fig:embedded-data}.
\end{itemize}

\begin{figure}[ht]
\centering
\begin{subfigure}{.5\textwidth}
  \centering
  \includegraphics[width=1.\linewidth]{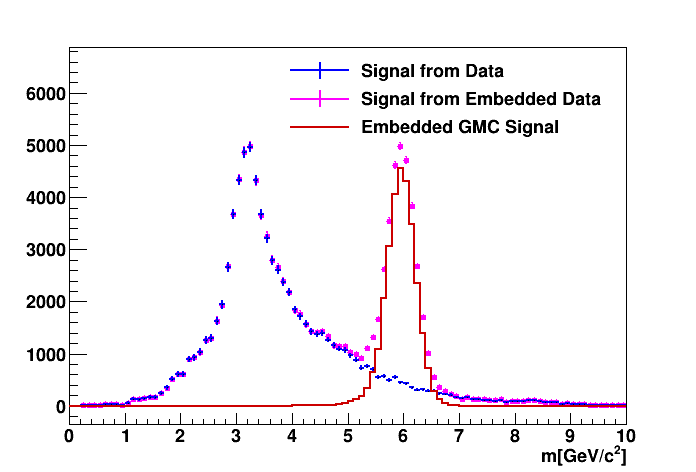}
\end{subfigure}%
\begin{subfigure}{.5\textwidth}
  \centering
  \includegraphics[width=1.\linewidth]{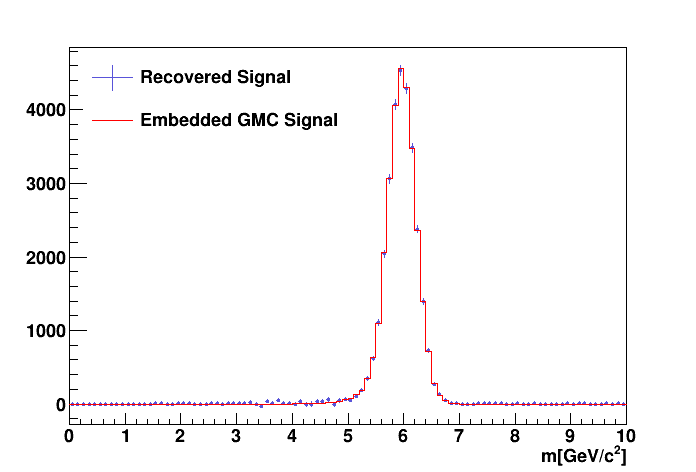}
\end{subfigure}
\caption{Left: Comparison of signals from different stages of analysis. The blue is total signal (Signal1) from the 10 normal runs, magenta is signal from simulated embedded data (Signal2); both are obtained after subtracting the combinatoric background. The red histogram is the embedded simulated signal, same as Fig.~\ref{fig:embed_signal}. Right: The blue histogram is the difference (Signal2 -- Signal1) between the blue and magenta histograms in the left-hand panel.  The red is the embedded signal, same as Fig.~\ref{fig:embed_signal}.}
\label{fig:all_signals}
\end{figure}

\begin{figure}[ht]
\centering
\includegraphics[width=0.65\linewidth]{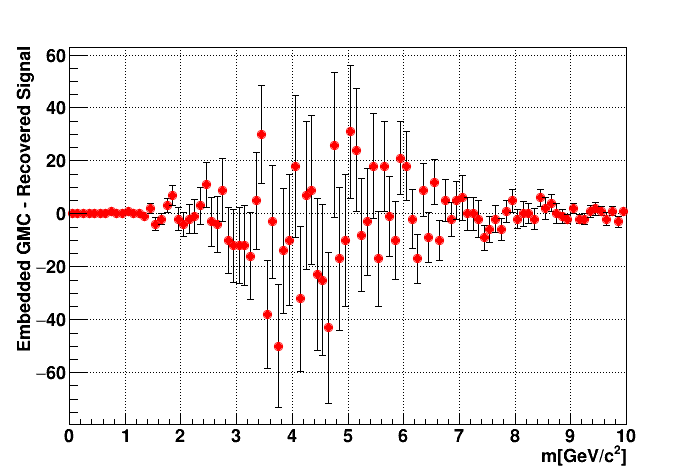}
\caption{The difference between signal recovered and signal embedded; this is the difference between the two histograms in Fig.~\ref{fig:all_signals} right, above.}
\label{fig:diff}
\end{figure}

The left hand side of Fig.~\ref{fig:all_signals} shows the various signals as we discussed above. The magenta points are the signal we obtained from the embedded data, the blue points show the signal from the real data, and the red histogram shows the total simulated signal we embedded. So, if the normalization from the mixing method is unity then the difference between the signal from embedded data (Signal2) and that from real data (Signal1) should be same as the embedded data. That difference is shown as the blue points in the right hand side of the Fig.~\ref{fig:all_signals} together with the red signal histogram. The figure shows that the two distributions are in good agreement.

 Figure~\ref{fig:diff} shows the difference between the embedded simulated signal and signal recovered from the mixing method. The distribution has statistical fluctuations centered around zero; there is no residual signal.  

\begin{figure}[ht]
\centering
\includegraphics[width=.49\linewidth]{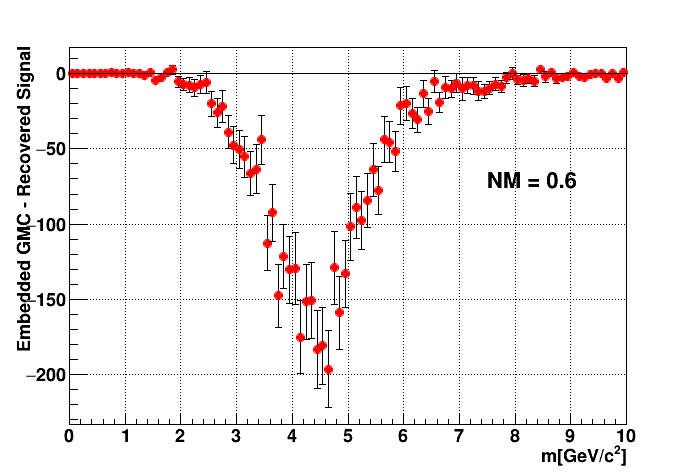}
\includegraphics[width=.49\linewidth]{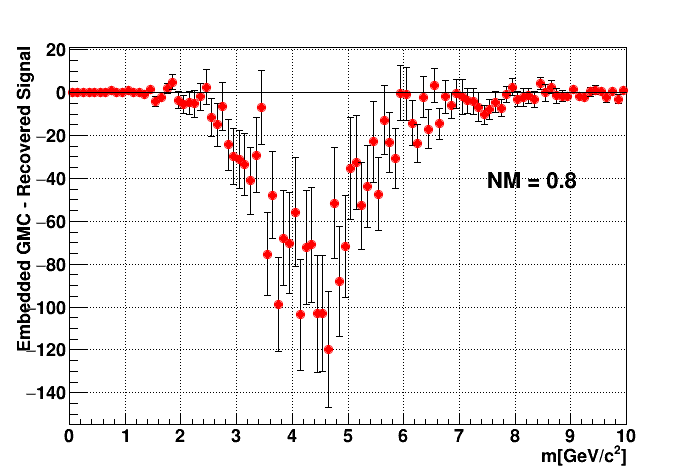}
\includegraphics[width=.49\linewidth]{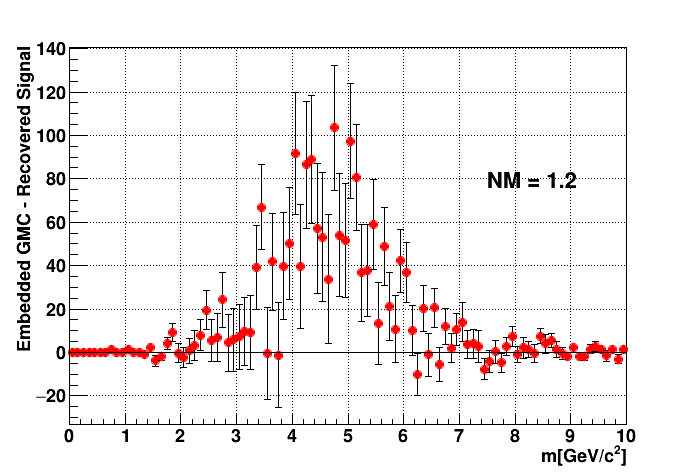}
\includegraphics[width=.49\linewidth]{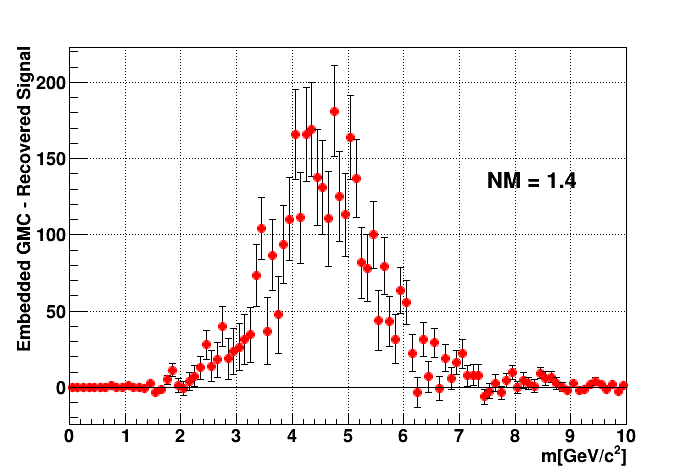}
\caption{The difference between signal recovered and signal embedded when both mixed distributions as described in Sec.~\ref{sec:cons_check} are scaled by a common normalization factor. Compare to Fig.~\ref{fig:diff} where NM=1.}
\label{fig:Peng_test}
\end{figure}

As an additional test of our claim of correct normalization, we attach a common normalization factor (NM) to the combinatoric backgrounds and see if the residual spectrum (like Fig.~\ref{fig:diff}) is affected.  This means we will calculate
\begin{center}
Signal1 = Yield1 -- NM*Comb1 ~~~~~ and ~~~~~ Signal2 = Yield2 -- NM*Comb2.
\end{center}
The result is shown in Fig.~\ref{fig:Peng_test}.  We see that any choice of NM other than 1 leaves a residual signal.  This is numerically demonstrated in Table~\ref{tab:nm_tab_1}, where we have integrated the area of each histogram in Figs.~\ref{fig:diff} and \ref{fig:Peng_test}.

A number of other tests, of Type 1 and Type 2, are shown in the Appendix.  In all of these tests, the embedded events have been placed at fixed intervals, so that the adjacent signals term $N_{\rm AS}$ has been turned off.  We show in all these cases that the $N_C'$ term generated in the mixed data correctly estimates the combinatoric background $N_C$.  In the next Section we turn to the issue of the effect of adjacent signals in the event stream.
\begin{table}[ht]
\begin{center}
\begin{tabular}{c c c c c c}
Normalization & $0.6$ & $0.8$ & $1.0$ & $1.2$ & $1.4$ \\
\hline
Area      & $-3610\pm 118$ & $-1876\pm 125$ & $-143\pm 132$ & $1590\pm 138$ & $3324\pm 145$ \\
\end{tabular}
\end{center}
\caption{\label{tab:nm_tab_1} Area of each histogram in Figs.~\ref{fig:diff} and \ref{fig:Peng_test} as a function of the normalization factor NM.}
\end{table}

\section{Quantification of the Effect of the Adjacent Signals Term}
\label{sec:maxSB}

Previously, in Sec.~\ref{sec:norm}, we mentioned that the adjacent signals term might not be negligible under some circumstances.  This does not invalidate the use of the proposed mixing method, because the size and effect of this term can be quantified using the simulation and embedding methods described in Sec.~\ref{sec:cons_check}.  In this section, we begin to embed the simulated signals randomly into the event stream, thus introducing the possibility of signals being in adjacent events.

We use a Type 1 test as described above in Sec.~\ref{sec:cons_check}. We show the results for each of the 10 runs, and we show results both for embedding the simulated signals at fixed intervals and at random locations in the event stream.  The simulated signals were embedded into 5\% of the events.  In Table~\ref{tab:fixedvsrandom} we show the residual signal after subtracting the recovered signal from the generated signal.  The residual signal is the integral of the difference.  In each run, when embedding at fixed intervals, the residual signal statistically fluctuates around zero.  On the other hand, when embedding randomly, the effect of the adjacent signals term is seen as a net positive residual signal.

In each of the ten runs, there are about 150,000 events.  In this test, we embedded simulated signal pairs into 5\% of those events, that is about 7500 embedded signal pairs.   When we embed randomly in the event stream, then about 0.25\% of events will have embedded signals in adjacent events, that is about 375 such events; this would be the value of the $N_{\rm AS}$ term in the mixed run.  However, not all of those 375 mixed pairs will pass the top/bottom-trigger condition, nor will all of them form a proper dimuon vertex, because they are a pair of unrelated tracks.  So, we should see fewer than $N_{\rm AS}$ extra counts in our residual signal.  Table~\ref{tab:fixedvsrandom} confirms this; the residual signal is $192.6\pm 41.5$ and not 375.  The $N_{\rm AS}$ term is an upper limit on the size of the final residual signal produced by adjacent signal pairs; additional analysis requirements (``cuts'') will reduce the effect of these pairs.

\begin{table}[ht]
\begin{center}
\begin{tabular}{c|c|c}
\hline
Run Number & Fixed Embedding & Random Embedding \\ 
\hline
12525 & $143\pm 137.7$ & $90\pm 138.1$ \\
12527 & $-43\pm 146.4$ & $206\pm 147.3$ \\
12528 & $1\pm 143.3$ & $225\pm 144.4$ \\
12529 & $-14\pm 109.6$ & $148\pm 109.9$ \\
12530 & $84\pm 139.4$ & $188\pm 140.4$ \\
12531 & $-22\pm 147.5$ & $42\pm 148.1$ \\
12532 & $37\pm 91.3$ & $160\pm 92.3$ \\
12533 & $-174\pm 146.8$ & $119\pm 147.6$ \\
12534 & $150\pm 149.7$ & $487\pm 150.1$ \\
12535 & $182\pm 145.0$ & $364\pm 145.2$ \\ 
\hline
Weighted Average & $33.1\pm 41.3$ & $192.6\pm 41.5$ \\
\hline
\end{tabular}
\caption{List of residual signals in each run, for embedding done at fixed intervals in the event stream and for embedding done randomly, using a Type 1 test.}
\label{tab:fixedvsrandom}
\end{center}
\end{table}

\section{Conclusion}
We have developed a method to estimate the combinatoric background valid for dilepton experiments where (1) the population density of signal pairs in the data stream is sufficiently low, and (2) the events can be sorted into classes containing the same track distributions.  The method has the correct normalization and the computed distribution can be directly subtracted from the total yields to recover the signal yields.  In the case of experiments with sufficiently high statistical significance, the effect of signal pairs that occur in adjacent events can perturb the results, and we have demonstrated a technique for quantifying this effect and correcting for it.

In an experiment with low statistics for both the signal and the background, it can be desirable to improve the statistical significance of the estimate of the background.  In principle, one could double the statistics of the estimated combinatoric background by combining the positive tracks from event $i$ with the negative tracks from both event $i+1$ and $i+2$ and retain the correct normalization by dividing by 2.  (And one could imagine extending this to events $i+3$, $i+4$, etc.)  We have not explored this.  The effect of the adjacent signals would need to be investigated for this case; doubling the size of the event pool will double the number of ``adjacent signals''.

Other groups attempting to use this method should perform the same sorts of tests, as we have shown here, to make sure this method is applicable to their situation.

\section{Acknowledgments}

This work was partially supported under grants DE-FG02-94ER40847 (New Mexico State), DE-AC02-06CH11357 (Argonne), DE-FG02-07ER41528 (Mississippi State), DE-FG02-96ER40950 (Virginia) from the US Department of Energy, Office of Nuclear Physics, as well as by the National Science Foundation under grants 2110229 \& 2012926 (Michigan) and 2013002 (Colorado).

\newpage
\appendix
\section{Appendix: Consistency Checks Performed by Embedding Various Simulated Distributions into Data }

Here we show additional tests we have done using the embedding of simulated tracks into real track data.  The embedding here is not done randomly, but instead at fixed intervals in the event stream, thus turning off the effect of the adjacent signals term $N_{\rm AS}$.

\subsection{Type 2 Test with 3.14 GeV/$c^2$ Dimuon Embedded Signal}
\begin{figure}[ht]
\centering
\includegraphics[width=0.5\linewidth]{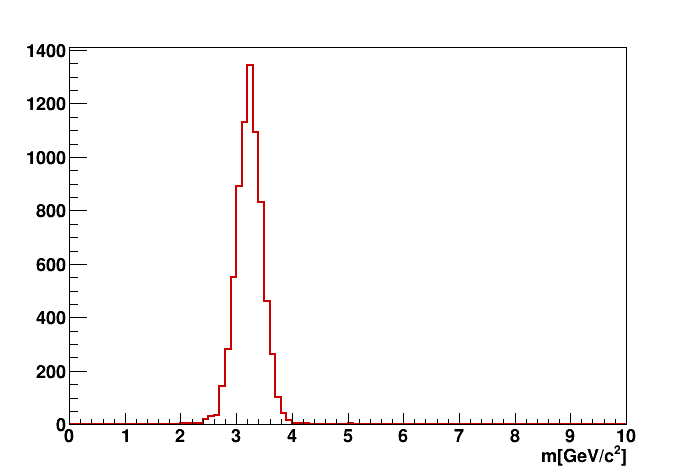}
\caption{The simulated 3.14-GeV/$c^2$ resonance signal that will be embedded into the data displayed in Fig.~\ref{fig:rs67-data}.}
\label{fig:mc_piGeV}
\end{figure}

\begin{figure}[ht]
\centering
\begin{subfigure}{.45\textwidth}
  \centering
  \includegraphics[width=1.\linewidth]{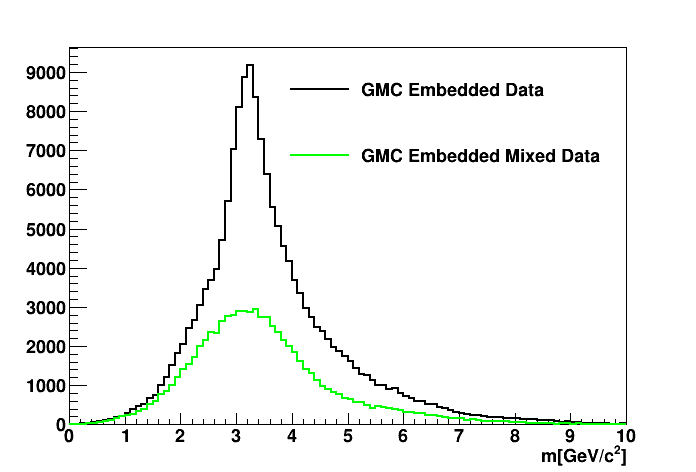}
\end{subfigure}%
\begin{subfigure}{.45\textwidth}
  \centering
  \includegraphics[width=1.\linewidth]{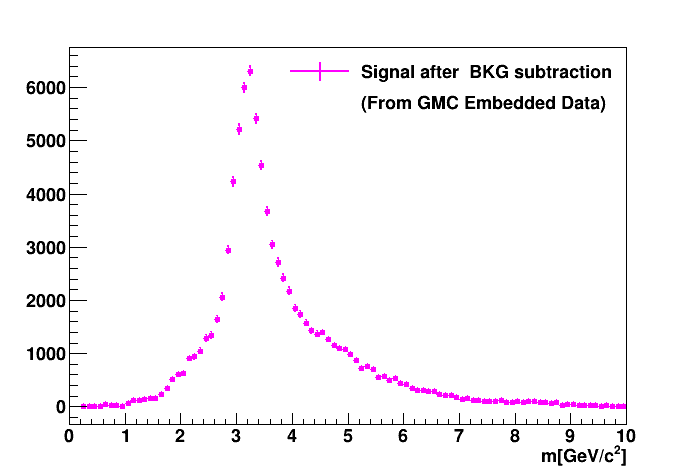}
\end{subfigure}
\caption{Left: Dimuon distribution (black) after simulated signal from Fig.~\ref{fig:mc_piGeV} is embedded into the data from Fig.~\ref{fig:rs67-data}, and the corresponding mixed distribution (green). Right: Signal obtained by subtracting mixed distribution from data.  Magenta = Black -- Green.}
\label{fig:embedded-data-piGeV}
\end{figure}

\begin{figure}[ht]
\centering
\centering
\begin{subfigure}{.45\textwidth}
  \centering
  \includegraphics[width=1.\linewidth]{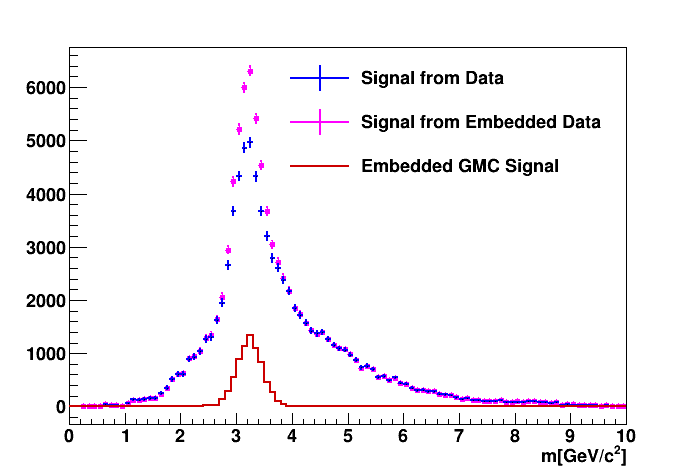}
\end{subfigure}%
\begin{subfigure}{.45\textwidth}
  \centering
  \includegraphics[width=1.\linewidth]{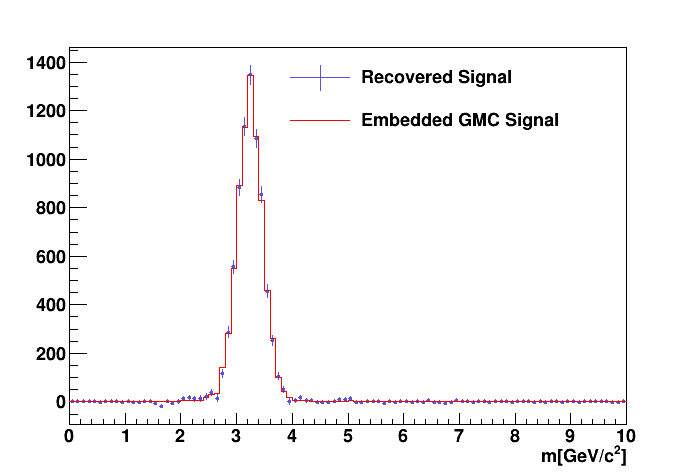}
\end{subfigure}
\caption{ Left: Comparison of signals from different stages of analysis. The blue points are the total signal from a single normal run; same as Fig.~\ref{fig:rs67-data} Right. The magenta points are the signal from data with the embedded simulated signal; same as Fig.~\ref{fig:embedded-data-piGeV} Right. Red is the embedded simulated signal, the same as Fig.~\ref{fig:mc_piGeV}. Right: Signal recovered (blue) vs.\ signal embedded (red). }
\label{fig:signal_comp_piGeV}
\end{figure}

\begin{figure}[ht]
\centering
\includegraphics[width=0.5\linewidth]{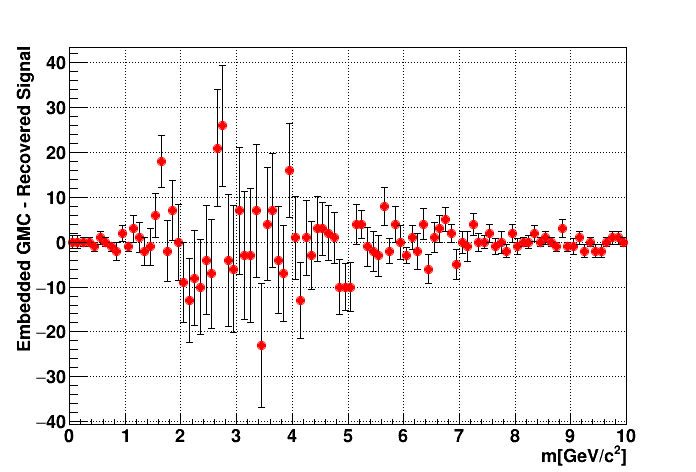}
\caption{The difference between signal recovered and signal embedded; this is the difference between the two histograms in the right hand side of Fig.~\ref{fig:signal_comp_piGeV} above. It is seen there is no residual signal.}
\label{fig:diff_piGeV}
\end{figure}

\begin{figure}[ht]
\centering
\includegraphics[width=.49\linewidth]{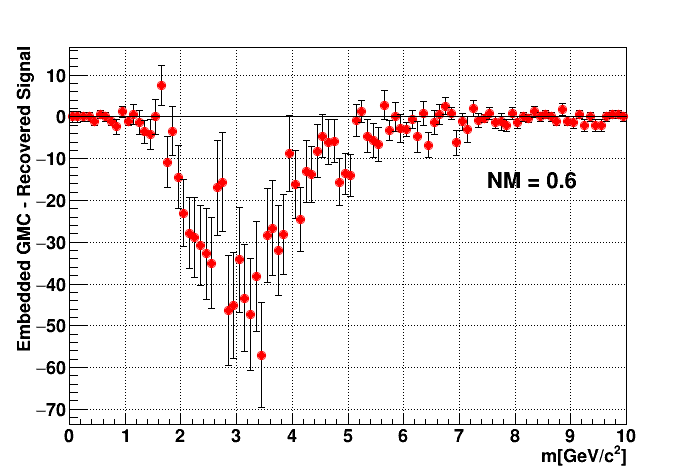}
\includegraphics[width=.49\linewidth]{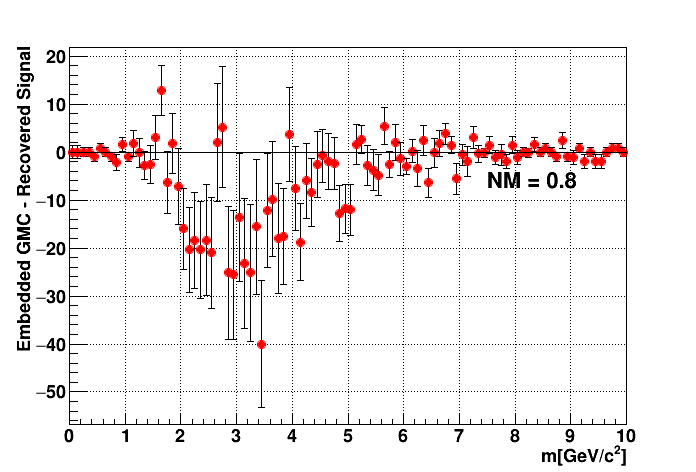}
.\includegraphics[width=.49\linewidth]{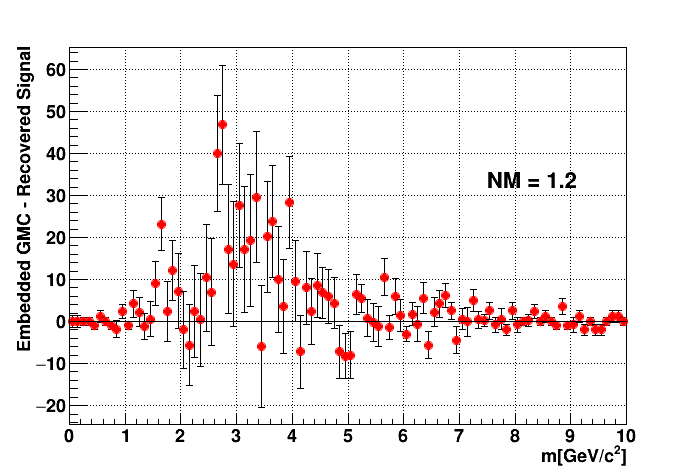}
\includegraphics[width=.49\linewidth]{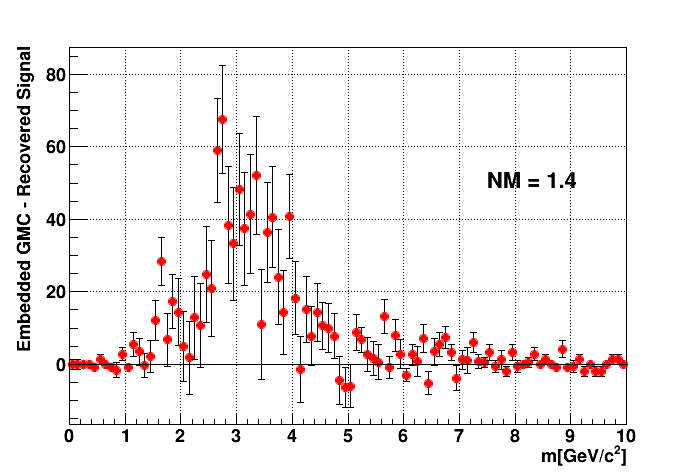}
\caption{The difference between signal recovered and signal embedded when both mixed distributions as described in Sec.~\ref{sec:cons_check} are scaled by different normalization factors. Compare to Fig.~\ref{fig:diff_piGeV} where NM=1.}
\label{fig:vary_nm_piGeV}
\end{figure}

Instead of a 6-GeV/$c^2$ resonance, we used a 3.14-GeV/$c^2$ resonance, because this places the embedded signal in the middle of the region with the largest yield. Here we embed the simulated signals in every 200$^{\rm th}$ event. We also included the variable normalization factor NM described above, and a table of the residual signals showing that the best choice is NM=1.  Please read the captions for Figs.~\ref{fig:mc_piGeV}-\ref{fig:vary_nm_piGeV} and see Table~\ref{tab:nm_tab_2}.

\clearpage

\clearpage

\begin{table}[ht]
\begin{center}
\begin{tabular}{c c c c c c}
Normalization & $0.6$ & $0.8$ & $1.0$ & $1.2$ & $1.4$ \\
\hline
Area      & $-858\pm 58$ & $-430\pm 62$ & $-1\pm 65$ & $428\pm 68$ & $856\pm 72$ \\
\end{tabular}
\end{center}
\caption{\label{tab:nm_tab_2} Area of each histogram in Figs.~\ref{fig:diff_piGeV} and \ref{fig:vary_nm_piGeV} as a function of the normalization factor NM.}
\end{table}

\clearpage
\newpage

\subsection{Type 2 Test Embedding Simulated Drell-Yan Events into Real Data }
\label{Type2DY}

\begin{figure}[ht]
\centering
\begin{subfigure}{.5\textwidth}
  \centering
  \includegraphics[width=.9\linewidth]{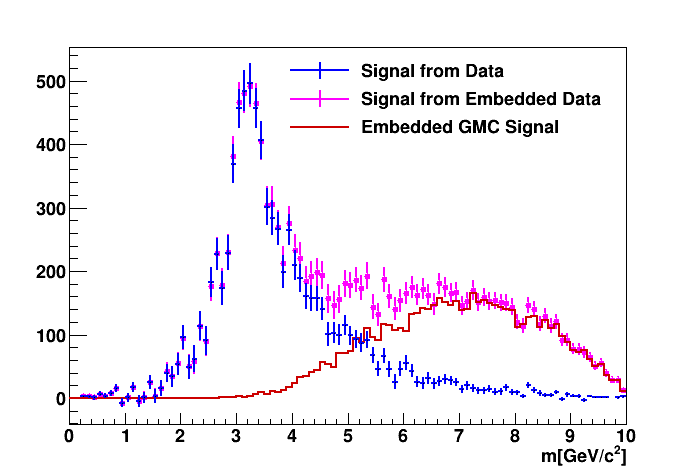}
\end{subfigure}%
\begin{subfigure}{.5\textwidth}
  \centering
  \includegraphics[width=.9\linewidth]{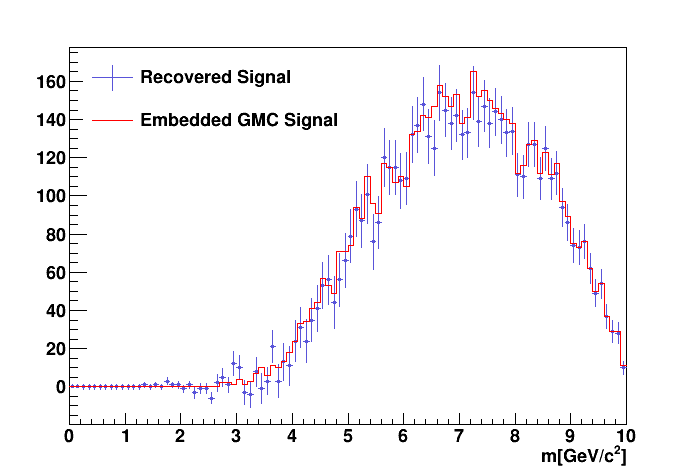}
\end{subfigure}
\caption{Left: Comparison of signals from different stages of analysis. The blue points are the total signal from a single normal data run, while the magenta points are the signal from data with a simulated embedded signal. Both are obtained after subtracting the combinatoric spectrum calculated from the event-mixing method. The red histogram is the embedded simulated signal. Right: Signal recovered (blue) vs. signal embedded (red).}
\label{fig:all_signals_unwt_DY}
\end{figure}

For this consistency check, we took one normal data run. We embedded reconstructed dimuon tracks from simulated Drell-Yan events in every 25$^{\rm th}$ event. The Drell-Yan events were generated uniformly in the mass range 0-10 GeV/$c^2$, and then passed through the detector simulation and reconstruction package.  After embedding, we implemented the mixing procedure to get the mixed distribution. Finally, the mixed distribution is subtracted from the embedded data distribution. The signal distribution thus recovered is consistent with the embedded signal. Please see Fig.~\ref{fig:all_signals_unwt_DY}.

\newpage

\subsection{Type 1 Test with 6-GeV/$c^2$ Dimuon Embedded Signal}
\begin{figure}[ht]
\centering
\includegraphics[width=.45\linewidth]{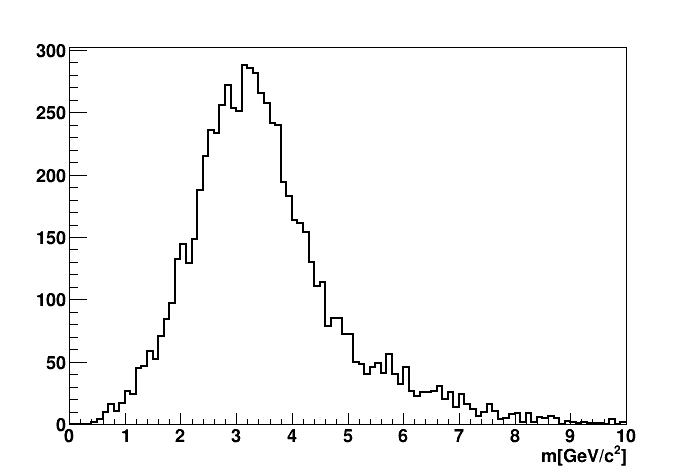}
\includegraphics[width=.45\linewidth]{fig/embed_signal-6GeV.png}
\caption{Left: Dimuon mass distribution from uncorrelated data (obtained from applying mixing method in real data). Right: Simulated signals to be embedded in uncorrelated data.}
\label{fig:a3_data_sim}
\end{figure}

\begin{figure}[ht]
\centering
\includegraphics[width=.45\linewidth]{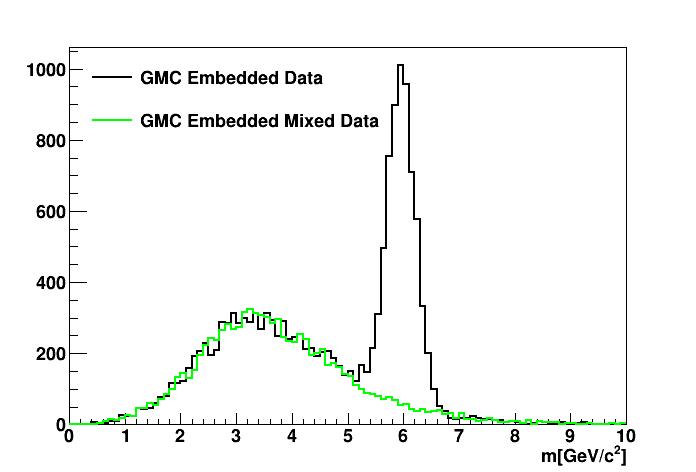}
\includegraphics[width=.45\linewidth]{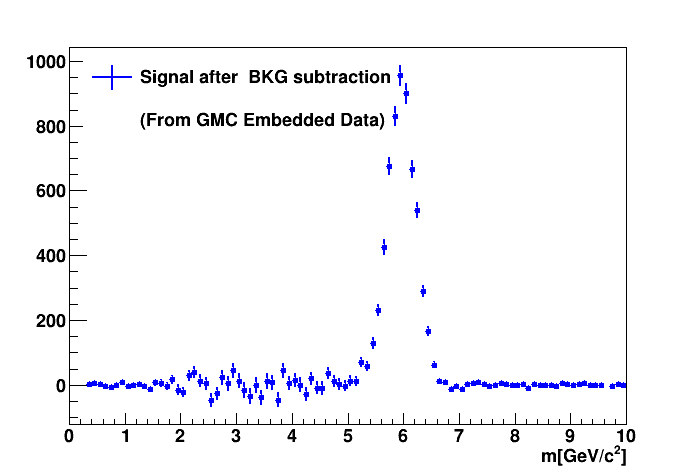}
\caption{Left: Simulated dimuon mass distribution embedded in uncorrelated data (black), and distribution from corresponding mixed events (green). Right: Signal obtained from subtracting.  Blue = Black -- Green.}
\label{fig:a3_signal_subtract}
\end{figure}

\begin{figure}[ht]
\centering
\includegraphics[width=.6\linewidth]{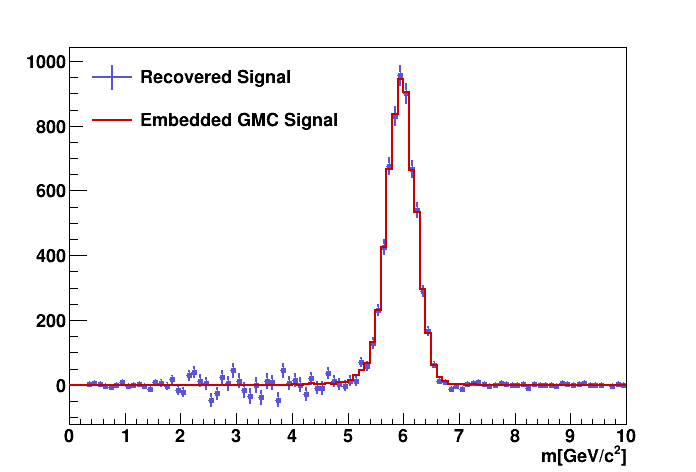}
\caption{Signal recovered from mixing method (blue points) and embedded simulated events (red line) plotted together.}
\label{fig:a3_compare}
\end{figure}

In this case, a simulated 6-GeV/$c^2$ resonance (Fig.~\ref{fig:a3_data_sim} right) is embedded into a mixed run (Fig.~\ref{fig:a3_data_sim} left), using every 25$^{\rm th}$ event.  The resulting total spectrum and combinatoric background (Fig.~\ref{fig:a3_signal_subtract} left) and the difference between them (Fig.~\ref{fig:a3_signal_subtract} right) are shown; the embedded signal and the extracted signal are compared directly in Fig.~\ref{fig:a3_compare}.

\clearpage

\subsection{Type 1 Test with Simulated Drell-Yan Dimuon Distribution}
\begin{figure}[ht]
\centering
\includegraphics[width=.45\linewidth]{fig/uncorData.png}
\includegraphics[width=.45\linewidth]{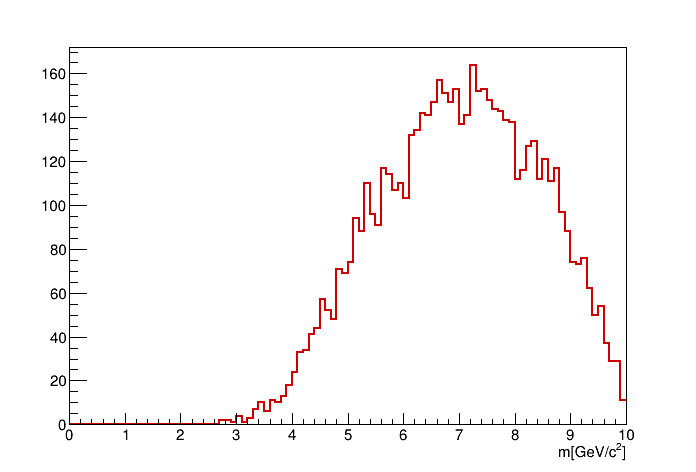}
\caption{Left: Dimuon mass distribution from uncorrelated data (obtained from applying mixing
method in real data), same as Fig.~\ref{fig:a3_data_sim}. Right: Simulated signals to be embedded in uncorrelated data.}
\label{fig:a4_data_sim}
\end{figure}

\begin{figure}[ht]
\centering
\includegraphics[width=.45\linewidth]{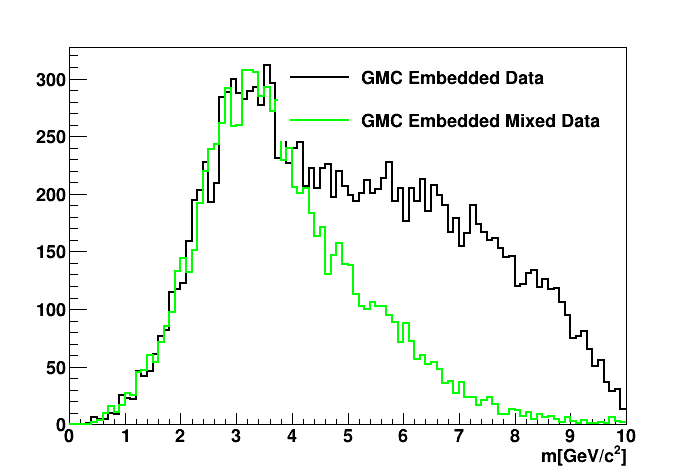}
\includegraphics[width=.45\linewidth]{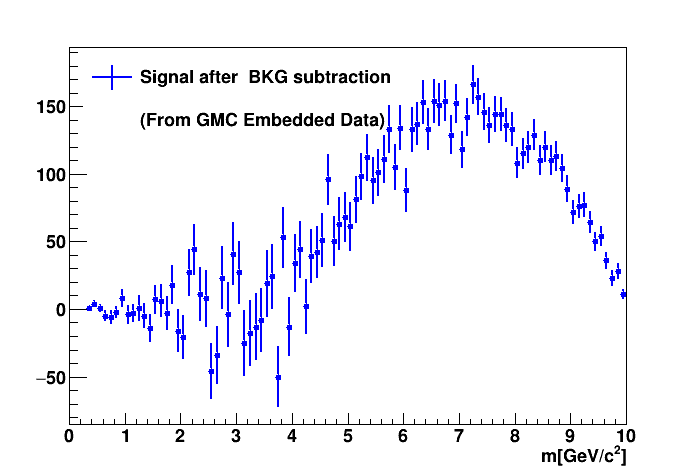}
\caption{ Left: Simulated dimuon mass distribution embedded into uncorrelated data (black) and distribution from corresponding mixed events (green). Right: Signal obtained from subtracting.  Blue = Black -- Green}
\label{fig:a4_signal_subtract}
\end{figure}

\begin{figure}[ht]
\centering
\includegraphics[width=.5\linewidth]{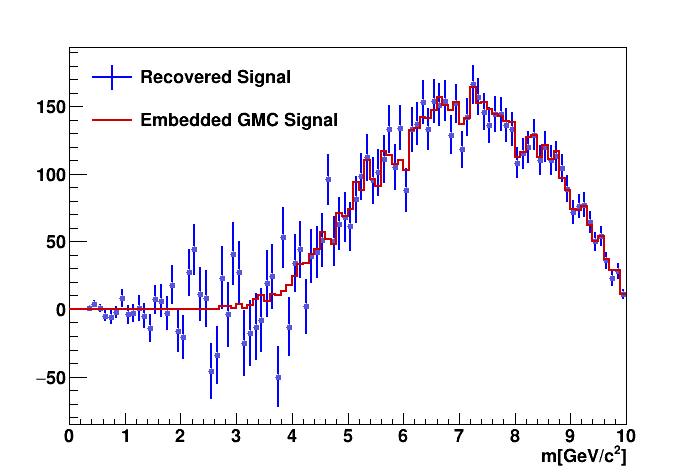}
\caption{Signal recovered from mixing method (blue points, Fig.~\ref{fig:a4_signal_subtract} right), and the embedded simulated events (red line, Fig.~\ref{fig:a4_data_sim} right), plotted together.}
\label{fig:a4_compare}
\end{figure}

In this case, a spectrum of simulated Drell-Yan events (Fig.~\ref{fig:a4_data_sim} right) is embedded into a mixed run (Fig.~\ref{fig:a4_data_sim} left), in every 25$^{\rm th}$ event.  This is the same spectrum of Drell-Yan events used in Subsection~\ref{Type2DY} above. The resulting total spectrum and combinatoric background (Fig.~\ref{fig:a4_signal_subtract} left) and the difference between them (Fig.~\ref{fig:a4_signal_subtract} right) are shown; the embedded signal and the extracted signal are compared directly in Fig.~\ref{fig:a4_compare}.

\clearpage

\bibliography{main}{}

\end{document}